\documentclass[final,3p,times,twocolumn]{elsarticle}
\usepackage{graphicx}
\usepackage{array}
\usepackage{algorithm2e}
\usepackage{caption}
\usepackage{amssymb}
\usepackage{amsmath}
\usepackage{bm}
\usepackage{amsthm}
\usepackage{dirtree}
\usepackage{lmodern}
\usepackage{subfigure}
\usepackage{hyperref}
\graphicspath{{./figs/}}
\usepackage{listings}
\newcommand{\belowtitleskip}{2pt}%\smallskipamount}

\usepackage{xcolor}
\definecolor{gray}{gray}{0.97}
\colorlet{commentcolour}{green!50!black}
\colorlet{stringcolour}{red!60!black}
\colorlet{keywordcolour}{magenta!90!black}
\colorlet{exceptioncolour}{yellow!50!red}
\colorlet{commandcolour}{blue!60!black}
\colorlet{promptcolour}{green!50!black}

\lstdefinestyle{pythonstyle}{
keepspaces=true,
language=python,
showtabs=true,
tab=,
tabsize=2,
basicstyle=\ttfamily\footnotesize,%\setstretch{.5},
stringstyle=\color{stringcolour},
showstringspaces=false,
alsoletter={1234567890},
otherkeywords={\ , \}, \{, \%, \&, \|},
keywordstyle=\color{keywordcolour}\bfseries,%
emph={[2]True, False, None},
emphstyle=[2]\color{keywordcolour},
emph={[3]object,type,isinstance,copy,deepcopy,zip,enumerate,
reversed,list,len,dict,tuple,xrange,append,execfile,real,imag,
reduce,str,repr, vars},
emphstyle=[3]\color{commandcolour},
emph={Exception,NameError,IndexError,SyntaxError,TypeError,ValueError,OverflowError,ZeroDivisionError},
emphstyle=\color{exceptioncolour}\bfseries,
morestring=[s]{"""}{"""},
morestring=[s]{'''}{'''},
commentstyle=\color{commentcolour}\slshape,
emph={[4]1, 2, 3, 4, 5, 6, 7, 8, 9, 0, 001, ode, fsolve, sqrt, exp, sin, cos, arccos, pi, array, norm, dot, arange, , isscalar, max, sum, flatten, shape, reshape, find, any, all, abs, plot, linspace, legend, quad, polyval,polyfit, hstack, concatenate,vstack,column_stack,empty,zeros,ones,rand,vander,
grid,pcolor,eig,eigs,eigvals,svd,qr,tan,det,logspace,roll,min,
mean,cumsum,cumprod,diff,vectorize,lstsq,cla,eye,xlabel,ylabel,
},
emphstyle=[4]\color{commandcolour},
emph={[5]and,break,class,continue,def,yield,del,elif ,else,
except,exec,finally,for,global,if,in,
lambda,not,or,pass,print,raise,return,try,while,assert},
emphstyle=[5]\color{black}\bf,
literate=*%
{>>>}{{\textcolor{promptcolour}{>>>}}}{1}%
,%
breaklines=true,
breakatwhitespace= true,
xleftmargin=0ex,
xrightmargin=0ex,
aboveskip=1ex,
belowskip=1ex,
frame=trbl, %trbl
numbers=none,
backgroundcolor=\color{white}
}

\lstdefinestyle{inlinestyle}{
language=python,
basicstyle=\ttfamily\footnotesize,%\setstretch{.5},
breaklines=true,
breakatwhitespace= true,
xleftmargin=0ex,
xrightmargin=0ex,
aboveskip=1ex,
belowskip=1ex,
frame=trbl, %trbl
numbers=none,
float=htpb,
}

\lstdefinestyle{scriptstyle}{
language=python,
numbers=left,
}

\lstnewenvironment{python}[1][]{
\lstset{style=pythonstyle, frame=trbl, belowcaptionskip=\belowtitleskip}
}{}

\lstnewenvironment{python_num}[1][]{
\lstset{style=pythonstyle, numbers=left, frame=trbl, 
belowcaptionskip=\belowtitleskip}
}{}

\newcommand{\inpyth}{\lstinline[style=inlinestyle]} %[]%

\newcounter{bla}

\journal{Computer Physics Communications}

\begin{document}

\begin{frontmatter}



\title{High performance Python for direct numerical simulations of turbulent flows}


\author[a,b]{Mikael Mortensen\corref{author}}
\author[b,c]{Hans Petter Langtangen}
\address[a]{Department of Mathematics, University of Oslo, Norway}
\address[b]{Center for Biomedical Computing at Simula Research Laboratory, N-1325 Lysaker, Norway}
\address[c]{Department of Informatics, University of Oslo, Norway}

\cortext[author] {Corresponding author.\\\textit{E-mail address:} mikaem@math.uio.no}

\newcommand{\hpl}[1]{({\bf hpl comment:} \emph{#1})}

\bibliographystyle{plain}

%\author{Mikael Mortensen\thanks{Department of Mathematics, University of Oslo, and Center for Biomedical Computing, Simula Research Laboratory. Email: \texttt{mikael.mortensen@gmail.com}.} and Hans Petter Langtangen\thanks{Center for Biomedical Computing, Simula Research Laboratory, and Department of Informatics, University of Oslo. Email: \texttt{hpl@simula.no}.}}
%\date{}							% Activate to display a given date or no date

\begin{abstract}
Direct Numerical Simulations (DNS) of the Navier Stokes equations is an
invaluable research tool in fluid dynamics. Still, there are few publicly
available research codes and, due to the heavy number crunching implied,
available codes are usually written in low-level
languages such as C/C++ or Fortran.
In this paper we describe a pure scientific Python pseudo-spectral DNS code
that nearly matches the performance of C++ for thousands of processors
and billions of unknowns. We also describe a version optimized through Cython, 
that is found to match the speed of C++. The solvers are written from scratch 
in Python, both the mesh, the MPI domain decomposition, and the temporal 
integrators. The solvers have been verified and benchmarked on the Shaheen 
supercomputer at the KAUST supercomputing laboratory, and we are able to show 
very good scaling up to several thousand cores.

A very important part of the implementation is the mesh decomposition (we 
implement both slab and pencil decompositions) and 3D parallel Fast Fourier 
Transforms (FFT). The mesh decomposition and FFT routines have been implemented 
in Python using serial FFT routines (either NumPy, pyFFTW or any other serial 
FFT module), NumPy array manipulations and with MPI communications handled by  
MPI for Python (\inpyth{mpi4py}). We show how we are able to execute a 3D 
parallel FFT in Python for a slab mesh decomposition using 4 lines of compact 
Python code, for which the parallel performance on Shaheen is found to be 
slightly better than similar routines provided through the FFTW library. For a 
pencil mesh decomposition 7 lines of code is required to execute a transform.

\end{abstract}


\end{frontmatter}

\section{Introduction}
%\subsection{}
Direct Numerical Simulations (DNS) is a term reserved for computer simulations 
of turbulent flows that are fully resolved in both time and space. DNS are 
usually conducted using numerical methods of such high order and accuracy that 
numerical dispersion and diffusion errors are negligible compared to their 
actual physical
counterparts. To this end, DNS has historically been carried out with extremely 
accurate and efficient spectral methods, and in the fluid dynamics community 
DNS enjoys  today the same status as carefully conducted experiments. DNS can 
provide detailed and highly reliable data not possible to extract from
experiments, which in recent years have driven a number of discoveries 
regarding the very nature of turbulence. The present paper presents a new, 
computationally attractive tool for performing DNS, realized by recent 
programming technologies.

Because of the extremely heavy number crunching implied by DNS,
researchers aim at highly optimized implementations running on
massively parallel computing platforms. The largest known DNS
simulations performed today are using hundreds of billions of degrees
of freedom, see, e.g., \cite{Lee2013, deBruynKops15}. Normally, this demands a 
need for developing tailored, hand-tuned
codes in what we here call low-level languages: Fortran, C or C++ (despite
the possibility for creating high-level abstractions in Fortran 90/2003 and
C++, the extreme performance demands of DNS codes naturally leads to
minimalistic use of classes and modules). Few DNS codes are openly available 
and easily accessible to the public and the common fluid mechanics researcher. 
Some exceptions are hit-3d (Fortran90) \cite{hit-3d}, Philofluid (Fortran) 
\cite{philofluid}, Tarang (C++) \cite{tarang}, and Turbo (Fortran90)
\cite{turbo}. However, the user interfaces to these codes are not sophisticated 
and user-friendly, and it is both challenging and time consuming for a user to
modify or extend the codes to satisfy their own needs. This is usually
the nature of codes written in low-level languages.

It is a clear trend in computational sciences over the last two decades
that researchers tend to move from low-level to high-level languages
like Matlab, Python, R, and IDL, where prototype solvers can be developed at 
greater comfort. The experience is that implementations
in high-level languages are faster to develop, easier to test,
easier to maintain, and they reach a much wider audience because the codes are 
compact and readable. The downside has been the decreased computational
efficiency of high-level languages and in particular their lack of
suitability for massively parallel computing. In a field like computational
fluid dynamics, this argument has been a show stopper. 

%Because of the massive amounts of number crunching involved, DNS solvers are usually implemented in high entry-level, low-level languages like Fortran/Fortran90 or C/C++.

Python is a high-level language that over the last two decades has grown very 
popular in the scientific computing community. A wide range of well 
established, ``gold standard'' scientific libraries in Fortran and C have been 
wrapped in Python, making them directly accessible just as commands in MATLAB. 
There is little overhead in calling low-level Fortran and C/C++ functions from 
Python, and the computational speed obtained in a few lines of code may easily 
compete with hundreds of compiled lines of Fortran or C code. It is important 
new knowledge in the CFD community if flow codes can be developed with comfort 
and ease in Python without sacrificing much computational efficiency.

The ability of Python to wrap low-level, computationally highly efficient 
Fortran and C/C++ libraries for various applications is today well known, 
appreciated, and utilized by many. A lesser known fact is that basic scientific 
Python modules like NumPy (cf. \cite{numpy, van2011numpy}), used for linear 
algebra and array manipulations, and MPI for Python (mpi4py) \cite{mpi4py}, 
which wraps (nearly) the entire MPI library, may be used directly to develop, 
from scratch, 
high performance solvers that run at speeds comparable to the very best 
implementations in low-level codes. A general misconception seems to be that 
Python may be used for fast prototyping and post-processing, as MATLAB, but 
that serious high-performance computing on parallel platforms require 
reimplementations in Fortran, C or C++. In this paper, we conquer this 
misconception: The only real requirement for developing a fast scientific 
Python solver is that all array manipulations are 
performed using vectorization (that calls underlying BLAS or LAPACK backends or 
compiled NumPy ufuncs) such that explicit for loops over long arrays in Python 
are avoided. The MPI for Python module in turn provides a message passing 
interface for NumPy arrays at communication speeds very close to pure 
C code.

There are already several examples on successful use of Python for
high-performance parallel scientific computing. The sophisticated
finite element framework FEniCS \cite{fenics} is written mainly in
C++, but most application developers are writing FEniCS-based solvers
directly in Python, never actually finding themselves in need of
writing longer C++ code and firing up a compiler. For large scale
applications the developed Python solvers are usually equally fast as
their C++ counterparts, because most of the computing time is
spent within the low-level wrapped C++ functions that perform the
costly linear algebra operations \cite{Mortensen2015}.
GPAW \cite{gpaw05} is a code devoted
to electronic structure calculations, written as a combination of
Python and C. GPAW solvers written in Python have been shown to scale
well for thousands of processors.  The PETSc project \cite{petsc-web-page} is a 
major provider of linear algebra to the open source community. PETSc was
developed in C, but through the package PETSc for Python (petsc4py) almost all
routines may be set up and called from Python. PyClaw \cite{ketcheson2012}
is another good example, providing a compact, powerful, and intuitive
Python interface to the algorithms within the Fortran codes Clawpack
and SharpClaw. PyClaw is parallelised through PETSc and has been shown
to scale well up to 65,000 cores.

Python has capabilities today for providing short and quick implementations 
that compete with tailored implementations in low-level languages up to 
thousands of processors. This fact is not well known, and the purpose of this 
paper is to demonstrate such a result for DNS and show the technical 
implementation details that are needed. As such, the major objective of this 
work is to explain a novel implementation of an excellent research tool (DNS) 
aimed at a wide audience. To this end, we i) show how a complete 
pseudo-spectral DNS solver can be written from scratch in Python using less 
than 100 lines of compact, very readable code, and ii) show that these 100 
lines of code can run at speeds comparable to its low-level counterpart in 
hand-written C++ code on thousands of processors. To establish scaling and 
benchmark results, we have run the codes on Shaheen, a massively parallel 
Blue Gene/P machine at the KAUST Supercomputing Laboratory. The code described 
is part of a larger DNS project and available online 
(https://github.com/mikaem/spectralDNS) under a GPL license. 

\section{The Navier-Stokes equations in spectral space}
Our DNS implementation is based on a pseudo-spectral Fourier-Galerkin method \cite{canuto1988} for the spatial discretization. The Navier-Stokes equations are first cast in rotational form

\begin{align}
 \frac{\partial \bm{u}}{\partial t} - \bm{u} \times \bm{\omega}   &= \nu \nabla^2 \bm{u} - \nabla{P}, \label{eq:NS} \\
 \nabla \cdot \bm{u} &= 0, \\
 \bm{u}(\bm{x}+2\pi \bm{e}^i, t) &= \bm{u}(\bm{x}, t), \quad \text{for }\, i=1,2,3,\\
 \bm{u}(\bm{x}, 0) &= \bm{u}_0(\bm{x})
\end{align}
where $\bm{u}(\bm{x}, t)$ is the velocity vector, $\bm{\omega}=\nabla \times \bm{u}$ the vorticity vector, $\bm{e}^i$ the Cartesian unit vectors, and the modified pressure $P=p+\bm{u}\cdot \bm{u}/2$, where $p$ is the regular pressure normalized by the constant density. The equations are periodic in all three spatial directions. If all three directions now are discretized uniformly in space using a structured computational mesh with $N$ points in each direction, the mesh points can be represented as\footnote{Different domains lengths and number of points in each direction are trivially implemented, and we use a uniform mesh here for simplicity.}
\begin{multline}
\bm{x} = (x, y, z) = \Big\{(x_i, y_j, z_k) = \left( \frac{2\pi i}{N}, 
\frac{2\pi j}{N}, \frac{2\pi k}{N} \right): \\ 
\, i,j,k \in 0,\ldots, N-1 \Big\}.
\label{eq:realmesh}
\end{multline}
In the spectral Galerkin method all variables must be transformed from the physical mesh $\bm{x}$ to a discrete and bounded Fourier wavenumber mesh. The three-dimensional wavenumber mesh may be represented as
\begin{multline}
\bm{k} = (k_x, k_y, k_z) = \Big\{(l, m, n): \\ \, l, m, n \in 
-\frac{N}{2}+1,\ldots, \frac{N}{2} \Big\}.
\label{eq:kmesh}
\end{multline}
The discrete Fourier transforms are used to move between real space $\bm{x}$ and spectral space $\bm{k}$. A component of the velocity vector (with similar notation for other field variables) is approximated in both real and spectral space as
\begin{align}
u(\bm{x}, t) &= \frac{1}{N^3}\sum_{\bm{k}} \hat{u}_{\bm{k}}(t) e^{\imath \bm{k}\cdot \bm{x}}, \label{eq:ffteq} \\
\hat{u}_{\bm{k}}(t) &= \sum_{\bm{x}} u(\bm{x}, t) e^{-\imath \bm{k}\cdot \bm{x}},\label{eq:iffteq}
\end{align}
where $\hat{u}_{\bm{k}}(t)$ is used to represent the Fourier coefficients, $\imath=\sqrt{-1}$ represents the imaginary unit, and $e^{\imath \bm{k}\cdot \bm{x}}$ represents the basis functions for the spectral Galerkin method. Equations (\ref{eq:ffteq}) and (\ref{eq:iffteq}) correspond, respectively, to the three-dimensional discrete Fourier transform, $\mathcal{F}$, and its inverse $\mathcal{F}^{-1}$. We use the notation
\begin{multline}
\hat{u}_{\bm{k}}(t) = \hat{{u}}(k_x, k_y, k_z, t) = \mathcal{F}({u}(\bm{x}, 
t)) 
\\ \left[= \mathcal{F}_{k_x} \left(\mathcal{F}_{k_y} \left( \mathcal{F}_{k_z} 
({u}) \right) \right) \right], \label{eq:fft}
\end{multline}
\begin{multline}
{u}(\bm{x}, t) = {u}(x, y, z, t) = \mathcal{F}^{-1}(\hat{{u}}_{\bm{k}}(t)) \\ 
\left[= 
\mathcal{F}^{-1}_{z}\left(\mathcal{F}^{-1}_{y}\left(\mathcal{F}^{-1}_{x}(\hat{{u}})\right)\right)\right],
 \label{eq:ifft}
\end{multline}
where the indices on $\mathcal{F}$ and $\mathcal{F}^{-1}$ within the square 
brackets are used to indicate direction of the transform. Both forward and 
inverse transforms involve three consecutive transformations, one for each 
periodic direction. The inverse transforms must be computed in the opposite 
order of the forward transforms, but otherwise the order is arbitrary. The 
first forward transform (here in the $z$-direction) is real and the remaining 
two are complex valued.

%For completeness, the first real transform in the $z$-direction is defined for all $k_z = -N/2+1, \ldots, N/2$ as
%\begin{align}
%\mathcal{F}_{k_z}(\bm{u}(x, y, z, t)) = \hat{\bm{u}}(x, y, k_z, t) &= \frac{2\pi}{N}\sum_{k=0}^{N-1}{\bm{u}(x, y, {z}_k, t)}e^{-\imath k_z z_k},
%\end{align}
%with the inverse transform
%\begin{align}
%\mathcal{F}^{-1}_{z} (\hat{\bm{u}}(x, y, k_z, t) =\bm{u}(x, y, z, t) &= \frac{1}{2\pi}\sum_{n=-N/2+1}^{N/2}\hat{\bm{u}}(x, y, n, t)e^{\imath n z}.
%\end{align}
%Here $\imath=\sqrt{-1}$ is used to represent the imaginary unit.

The spectral Galerkin method involves taking the inner product of the Navier-Stokes equations with the basis functions, $e^{\imath \bm{k} \cdot \bm{x}}$, which corresponds to performing a three-dimensional Fourier transform of the Navier-Stokes equations. If we also evaluate the analytical spatial derivatives in spectral space, we obtain a system of ordinary differential equation for $\hat{\bm{u}}_{\bm{k}}$, whereas the continuity equation reduces to an orthogonal inner product. All in all we obtain
\begin{align}
 \frac{d\hat{\bm{u}}_{\bm{k}}}{d t} - \widehat{( \bm{u} \times \bm{\omega})}_{\bm{k}} &= - \nu |\bm{k}|^2  \hat{\bm{u}}_{\bm{k}} - \imath \bm{k} \hat{P}_{\bm{k}}, \label{eq:NSf} \\
 \imath \bm{k} \cdot \hat{\bm{u}}_{\bm{k}} &= 0.
\end{align}
The pressure may be eliminated by taking the divergence of (\ref{eq:NS}), or equivalently by dotting the transformed (\ref{eq:NSf}) by $\imath \bm{k}$ and rearranging such that
\begin{equation}
\hat{P}_{\bm{k}} = - \frac{\imath\bm{k} \cdot \widehat{( \bm{u} \times \bm{\omega})}_{\bm{k}} }{|\bm{k}|^2}.
\label{eq:mod_pressure}
\end{equation}
Inserting for the pressure in (\ref{eq:NSf}), the final equation to solve for the transformed velocity vector in wavenumber space is thus
\begin{equation}
 \frac{d\hat{\bm{u}}_{\bm{k}}}{d t}  = \widehat{( \bm{u} \times \bm{\omega})}_{\bm{k}} - \nu |\bm{k}|^2  \hat{\bm{u}}_{\bm{k}} - \bm{k} \frac{\bm{k} \cdot \widehat{( \bm{u} \times \bm{\omega})}_{\bm{k}} }{|\bm{k}|^2}. \label{eq:NSfinal}
\end{equation}
Note that the transformed velocity components are coupled through the nonlinear convection term and the eliminated pressure.

The pseudo-spectral label arises from the treatment of the convective term, which is computed by first transforming the velocity and vorticity to physical space, performing the cross product, and then transforming the vector ${(\bm{u}  \times  \bm{\omega})}$  back to Fourier space. The operation requires 2 inverse transforms (velocity and vorticity) and 1 forward transform for each of the three vector components, 9 all together. This is the only operation that requires MPI communication and it is typically the most computationally extensive part of a pseudo-spectral DNS solver.

The time integration of (\ref{eq:NSfinal}) is performed using an explicit method. Canuto et~al. \cite{canuto1988} provides a good discussion on many different integrators for these equations, and in the Appendix we have used a fourth-order Runge-Kutta method. For the DNS solver described here an explicit integrator is reasonable because the time step by physical measures is required to be small in order to resolve all temporal scales in the flow. And by resolving these scales the integrator is usually well within stability limits as dictated by the numerical scheme. In other words, the CFL number will usually be required to be much lower than 1 in order to resolve the temporal scales.

\section{Implementation}

We have implemented the pseudo-spectral discretization of the Navier-Stokes equations, as described in the previous section, in high-level Python code. A complete solver is shown in the Appendix. It is strongly remarked that this Python code is not simply a wrapper of a low-level, high-performance \emph{flow solver} written originally in Fortran, C or C++. The entire code is implemented directly in Python: the mesh, the solution arrays, the MPI domain decomposition, and the time integrators. We are only making use of wrappers for serial FFT, something that is also done by the majority of low-level flow solvers anyway. The current Python implementation may, as such, be used as an easy to follow, working prototype for a complete low-level implementation in Fortran, C or C++.

The scientific Python solver makes extensive use of the NumPy and MPI for 
Python (mpi4py) 
packages, but if the pyFFTW  module has been installed, this module 
will be used to perform the FFT instead of NumPy. Note that the 
pyFFTW module is a wrapper for the FFTW library (www.fftw.org), 
whereas NumPy wraps FFTPACK (www.netlib.org/fftpack).  The 
program starts with importing the necessary modules and initializing 
MPI:\footnote{Note that the coding style `from numpy import *` is generally not 
recommended because of namespace issues. We are making an exception here for 
brevity and because we are creating a solver not intended for import elsewhere.}

\begin{python}
from numpy import *
from numpy.fft import fftfreq, fft, ifft, \
  irfft2, rfft2, irfftn, rfftn
from mpi4py import MPI

try:
    from pyfftw.interfaces.numpy_fft \ 
    import fft, ifft, irfft2, rfft2, \
    irfftn, rfftn
except ImportError:
    pass # Rely on numpy.fft routines

# Get some MPI     
comm = MPI.COMM_WORLD
num_processes = comm.Get_size()
rank = comm.Get_rank()         
\end{python}
Importing \texttt{MPI} from \texttt{mpi4py} initializes the MPI communicator. 
Two different strategies, slab and pencil, have been implemented for the MPI 
parallel domain decomposition. However, since communication only enters the 
solver through the FFTs (and post processing), there is very little difference 
between a serial code and a parallel one. Therefore, we first present a serial 
version of the code.

\subsection{Serial version of code}
The computational mesh is in physical space a structured uniform (periodic) 
cube $[0, 2\pi]^3$, where each direction is divided into $N$ uniform intervals, 
where $N=2^M$ for a positive integer $M$. Any different size of the box may be 
easily implemented through scaling of the governing equations. The mesh 
according to (\ref{eq:realmesh}) is represented in Python as

\begin{python}
# The assigned size of the mesh
M = 6       
# Actual number of nodes in each direction
N = 2**M    
# Physical size of computational box
L = 2*pi    
# The mesh
X = mgrid[:N, :N, :N].astype(float)*L/N
\end{python}
The matrix \inpyth{X} has dimensions \inpyth{(3, N, N, N)}. Since the Navier 
Stokes equations are solved in Fourier space, the physical space is only used 
to compute the convection plus to do post processing. The mesh \inpyth{X} is 
typically used for initialization and is otherwise not needed (and may 
therefore be deleted to save memory). In parallel mode, \inpyth{X} will be 
split up and divided between the processors.  Note that the solver may be 
operated in either single or double precision mode, and that \inpyth{float} in 
our code is a placeholder for either one of the NumPy datatypes 
\inpyth{float32} or \inpyth{float64} (single or double precision), depending on 
settings.

The velocity field to be transformed is real, and the discrete Fourier transform of a real sequence has the property that $\hat{\bm{u}}_k = \hat{\bm{u}}_{N-k}^*$, where $^*$ denotes the complex conjugate. As such, it is sufficient to use $N/2+1$ Fourier coefficients in the first transformed direction, leading to a transformed wavenumber mesh of size $(N/2+1)N^2$. The odd number of wavenumbers does not lead to any issues for the serial version of the code or for the slab decomposition. However, for the pencil decomposition it requires special attention, see Section \ref{pencil2D}.

In our code the real transform is taken in the $z$ direction and the 
wavenumbers $\bm{k}=(k_x, k_y, k_z)$ stored on the transformed mesh thus has 
ordering as used by the FFT routines provided by NumPy or pyFFTW:

\begin{align}
  \bm{k} = [&(0, \ldots, N/2-1, -N/2, -N/2+1, \ldots, -1), \notag \\
   &(0, \ldots, N/2-1, -N/2, -N/2+1, \ldots, -1),  \notag \\
  &(0, \ldots, N/2-1, N/2)].
\end{align}
A three-dimensional wavenumber mesh is implemented as

\noindent
\begin{minipage}[l]{\columnwidth}
\begin{python}
Nf = N/2+1
kx = ky = fftfreq(N, 1./N).astype(int)
kz = kx[:Nf].copy(); kz[-1] *= -1
K = array(meshgrid(kx, ky, kz, 
          indexing='ij'), dtype=int)
K2 = sum(K*K, 0, dtype=int)
K_over_K2 = K.astype(float) / where(
            K2 == 0, 1, K2).astype(float)
\end{python}
\end{minipage}
where \inpyth{fftfreq(N, 1./N)} is a function that creates the wavenumbers $(0, 
\ldots, N/2-1, -N/2, -N/2+1, \ldots, -1)$. The dimensions of the matrices are 
\inpyth{(3, N, N, N/2+1)} for \texttt{K} and \texttt{K\_over\_K2}, and 
\inpyth{(N, N, N/2+1)} for \texttt{K2}, and these matrices represent $\bm{k}$, 
$\bm{k}/|\bm{k}|^2$ and $|\bm{k}|^2$, respectively. The last two matrices are 
precomputed for efficiency.

The velocity, curl and pressure are similarily stored in structured 
(uninitialized) NumPy arrays

\begin{python}
U     = empty((3, N, N, N),  dtype=float)
U_hat = empty((3, N, N, Nf), dtype=complex)
P     = empty((N, N, N),     dtype=float)
P_hat = empty((N, N, Nf),    dtype=complex)
curl  = empty((3, N, N, N),  dtype=float)
\end{python}
Here \inpyth{hat} denotes a transformed variable. To transform between, e.g., 
\inpyth{U} and \inpyth{U_hat}, calls to FFT routines are required. The three 
dimensional FFT and its inverse are implemented in Python functions as shown 
below.

\begin{python}
def fftn_mpi(u, fu):
    """FFT of u in three directions."""
    if num_processes == 1:   
        fu[:] = rfftn(u, axes=(0,1,2))
    return fu

def ifftn_mpi(fu, u):
    """Inverse FFT of fu in three directions."""
    if num_processes == 1:
        u[:] = irfftn(fu, axes=(0,1,2))
    return u

# Usage
U_hat = fftn_mpi(U, U_hat)
U = ifftn_mpi(U_hat, U)
\end{python}
For high performance, it is key that the Python code relies on \emph{in-place}
modifications of pre-allocated arrays to avoid unnecessary allocation of
large temporary arrays (which often arises from NumPy code with basic array 
arithmetics). 
Each of the functions above takes the result array (\texttt{U\_hat} or
\texttt{U}) as argument, fill this array with values and returns the
array to the calling code. A commonly applied convention in
Python is to return all result objects from functions as this only involves
transfer of references and no copying of data.

We also remark that the three consecutive transforms performed by 
\inpyth{rfftn/irfftn} are actually using one real transform along the 
$z$-direction and two complex transforms in the remaining two directions. Also 
note that the simple NumPy/pyFFTW wrapped functions \inpyth{rfftn/irfftn} for 
three dimensional FFT may only be used in single processor mode, and the MPI 
implementation is detailed in Sections
\ref{slab1D} and \ref{pencil2D}.

The convection term requires a transform from Fourier to physical space where the cross product $\bm{u} \times \bm{\omega}$ is carried out. The curl in Fourier space is
\begin{equation}
\mathcal{F}(\nabla \times \bm{u}) = \hat{\bm{\omega}}_{\bm{k}} = \imath \bm{k} \times \hat{\bm{u}}_{\bm{k}}.
\end{equation}
We can now compute the curl in physical space through $\bm{\omega} = \mathcal{F}^{-1}(\hat{\bm{\omega}}_{\bm{k}})$. The convection term may thus be computed as
\begin{equation}
\widehat{( \bm{u} \times \bm{\omega})}_{\bm{k}} = \mathcal{F}(\bm{u} \times \bm{\omega}) = \mathcal{F} (\bm{u} \times \mathcal{F}^{-1}(\imath \bm{k} \times \hat{\bm{u}}_{\bm{k}})).
\label{eq:curl_convection}
\end{equation}
The Python functions for the curl and cross products are

\begin{python}
def Cross(a, b, c):
    c[0] = fftn_mpi(a[1]*b[2]
                   -a[2]*b[1], c[0])
    c[1] = fftn_mpi(a[2]*b[0]
                   -a[0]*b[2], c[1])
    c[2] = fftn_mpi(a[0]*b[1]
                   -a[1]*b[0], c[2])
    return c

def Curl(a, c):
    c[2] = ifftn_mpi(1j*(K[0]*a[1]
                        -K[1]*a[0]), c[2])
    c[1] = ifftn_mpi(1j*(K[2]*a[0]
                        -K[0]*a[2]), c[1])
    c[0] = ifftn_mpi(1j*(K[1]*a[2]
                        -K[2]*a[1]), c[0])
    return c

\end{python}
and the computation of $\widehat{( \bm{u} \times \bm{\omega})}_{\bm{k}}$ is 
shown below in the Python function \inpyth{computeRHS}. The main body of the 
solver, including a simple Forward Euler integrator, is implemented as:

\begin{python}
# Array dU holds the right-hand side
dU = empty((3, N, Nf, N), dtype=complex)  
dt = 0.01    # Time step
nu = 0.001   # Viscosity

def computeRHS(dU):

    # Compute convective term
    curl = Curl(U_hat, curl)
    dU = Cross(U, curl, dU)

    # Dealias the nonlinear convection
    dU *= dealias

    # Compute pressure
    P_hat[:] = sum(dU*K_over_K2, 0, 
                   out=P_hat)

    # Subtract pressure gradient
    dU -= P_hat*K

    # Subtract viscous term
    dU -= nu*K2*U_hat

    return dU

t = 0        # Physical time
T = 1.0      # End time
while t <= T:
    t += dt
    U_hat += computeRHS(dU)*dt
    for i in range(3):
        U[i] = ifftn_mpi(U_hat[i], U[i])

\end{python}
Obviously, more advanced and accurate integrators may be easily added by modifying the final \inpyth{while} loop. See the Appendix for a fourth order Runge Kutta implementation.

The \inpyth{dealias} matrix multiplying \inpyth{dU} in the above
loop deserves a comment.
The nonlinear convection term in (\ref{eq:NSfinal}) is here dealiased using the 2/3-rule \cite{orzag71} and dealiasing is then simply achieved by an elementwise multiplication of a convection matrix \inpyth{dU[3, N, N, N/2+1]} with a matrix \inpyth{dealias[N, N, N/2+1]}, that is zero where the wavenumbers are larger than 2/3 of the Nyquist mode and unity otherwise. The relevant code to create such a matrix for dealiasing reads

\begin{python}
kmax_dealias = N/3.
dealias = array((abs(K[0]) < kmax_dealias)*
                (abs(K[1]) < kmax_dealias)*
                (abs(K[2]) < kmax_dealias), 
                dtype=bool)
\end{python}
Note that the dimensions of \inpyth{dU} and \inpyth{dealias} differ in the 
first index since \inpyth{dU} contains contributions from all three vector 
components. However, through automatic broadcasting, NumPy realizes that the 
last three dimensions are the same and as such all three components of 
\inpyth{dU} (i.e.,  \inpyth{dU[0]}, \inpyth{dU[1]} and  \inpyth{dU[2]}) are 
multiplied elementwise with the same matrix \inpyth{dealias}. Note that we have 
here simply described the most straightforward way of dealiasing a 
pseudo-spectral solver and that several more advanced and accurate methods 
exists, see, e.g., Canuto et~al. \cite{canuto1988}.

\subsection{1D Slab decomposition}
\label{slab1D}

To run the code on several processors using MPI, the mesh and the data 
structures need to be split up. The most popular strategy in the literature is 
the ``slab'' decomposition, where each CPU is assigned responsibility for a 
certain number of complete 2D planes (slices). In other words, just one of the 
three indices $(i,j,k)$ is split up and divided amongst the CPUs. The major 
drawback of the slab decomposition strategy is that the number of CPUs must be 
smaller than or equal to $N$ for a cubic mesh of size $N^3$. The MPI 
implementation in FFTW makes use of the slab decomposition strategy, but there 
is currently no interface from Python to these MPI routines.

The decomposed physical and wavenumber meshes for the slab decomposition are implemented through a slight modification of the serial code:

\begin{python}
Np = N / num_processes
sx = slice(rank*Np, (rank+1)*Np)
X = mgrid[sx, :N, :N].astype(float)*L/N
K = array(meshgrid(kx, ky[sx], kz,
          indexing='ij'), dtype=int)
\end{python}
In general, using \inpyth{num_processes} CPUs, each CPU gets responsibility for \inpyth{Np = N/num_processes} two-dimensional slices and the physical mesh is simply decomposed along the first index. The wavenumber mesh, on the other hand, is split along the second index. The reason for this choice is that the $k_x$ direction is the last direction to be transformed by the three consecutive FFTs and for this operation the data structures will need to be aligned in planes normal to the $k_y$ direction. This becomes obvious if one considers the MPI decomposition as illustrated in Figure~\ref{fig:Slabdecomp} for the case of a physical mesh of size $8^3$ divided amongst 4 CPUs.

The entire 3D parallel FFT may be implemented with preallocation of a work 
array for MPI and 4 lines (the body of \inpyth{fftn_mpi/ifftn_mpi}) of compact 
Python code, see Figure \ref{fig:fftn_slab}. Also shown in Figure 
\ref{fig:fftn_slab} is the parallel inverse transform.
\begin{figure*}
\begin{python}
# Preallocated work array for MPI
U_mpi = empty((num_processes, Np, Np, Nf), dtype=complex)

def fftn_mpi(u, fu):
    """FFT in three directions using MPI and the slab decomposition"""
    Uc_hatT = rfft2(u, axes=(1, 2))
    U_mpi[:] = rollaxis(Uc_hatT.reshape(Np, num_processes, Np, Nf), 1)
    comm.Alltoall([U_mpi, mpitype], [fu, mpitype])
    fu[:] = fft(fu, axis=0)
    return fu

def ifftn_mpi(fu, u):
    """Inverse FFT in three directions using MPI.
       Need to do ifft in reversed order of fft."""
    Uc_hat = ifft(fu, axis=0)
    comm.Alltoall([Uc_hat, mpitype], [U_mpi, mpitype])
    Uc_hatT[:] = rollaxis(U_mpi, 1).reshape(Uc_hatT.shape)
    u[:] = irfft2(Uc_hatT, axes=(2, 1))
    return u
\end{python}
\caption{Three dimensional forward \inpyth{fftn_mpi} and inverse \inpyth{ifftn_mpi} FFTs implemented for the slab decomposition.}
\label{fig:fftn_slab}
\end{figure*}
Consider the function \inpyth{fftn_mpi} in Figure \ref{fig:fftn_slab}. In the 
first call to \inpyth{rfft2} each processor performs a
complete two dimensional FFT in both $z$ and $y$ directions on the original real data structure \inpyth{u}
(i.e., the inner $\mathcal{F}_{k_y}(\mathcal{F}_{k_z}(u))$ from (\ref{eq:fft})).
The first transform, in the $z$-direction, is real to complex and the second complex to complex. The real
to complex transform reduces the size of the data structure to the one seen in Figure~\ref{slabsubfig1}.
To be able to perform the final transform in the $x$-direction, data must be communicated between all
processors. The second index of the data structure ($y$-direction, see Figure~\ref{slabsubfig2}) will now be
shared amongst the processors. The line involving \inpyth{rollaxis} is an 
advanced transpose operation that performs
the transformation from the data structure in Figure~\ref{slabsubfig1} to the one in Figure~\ref{slabsubfig2}.
After the transformation the data structures have the correct shape, but they contain the wrong data. The
communication of data required for the final transform takes place in one single MPI \inpyth{Alltoall}
operation, where the \inpyth{mpitype} is a placeholder for either \inpyth{MPI.F_DOUBLE_COMPLEX} or
\inpyth{MPI.F_FLOAT_COMPLEX} depending on settings. After this communication the data structures are lined
up for the final FFT in the $x$-direction. The \inpyth{ifftn_mpi} routine is basically the inverse of the
forward transform and its implementation should be straightforward to follow.

Only one explicit global pre-allocated complex work array 
(\inpyth{U_mpi[num_processes, Np, Np, N/2+1]}) is needed for the 
\inpyth{Alltoall} routine in the three-dimensional FFT. The transform routines 
of pyFFTW maintain their own work arrays and \inpyth{Uc_hatT} and 
\inpyth{Uc_hat} above actually point to this internal data storage. 
Alternatively, the forward data transfer may be 
performed completely without explicit work arrays using the in-place 
\inpyth{Sendrecv_replace} instead of \inpyth{Alltoall}, along with transpose 
operations. This is easily implemented, but, since the method has been found 
slightly slower, details are not shown here.

\begin{figure*}[th!]
\centering
\subfigure[Physical mesh]{
  \includegraphics[scale=0.15]{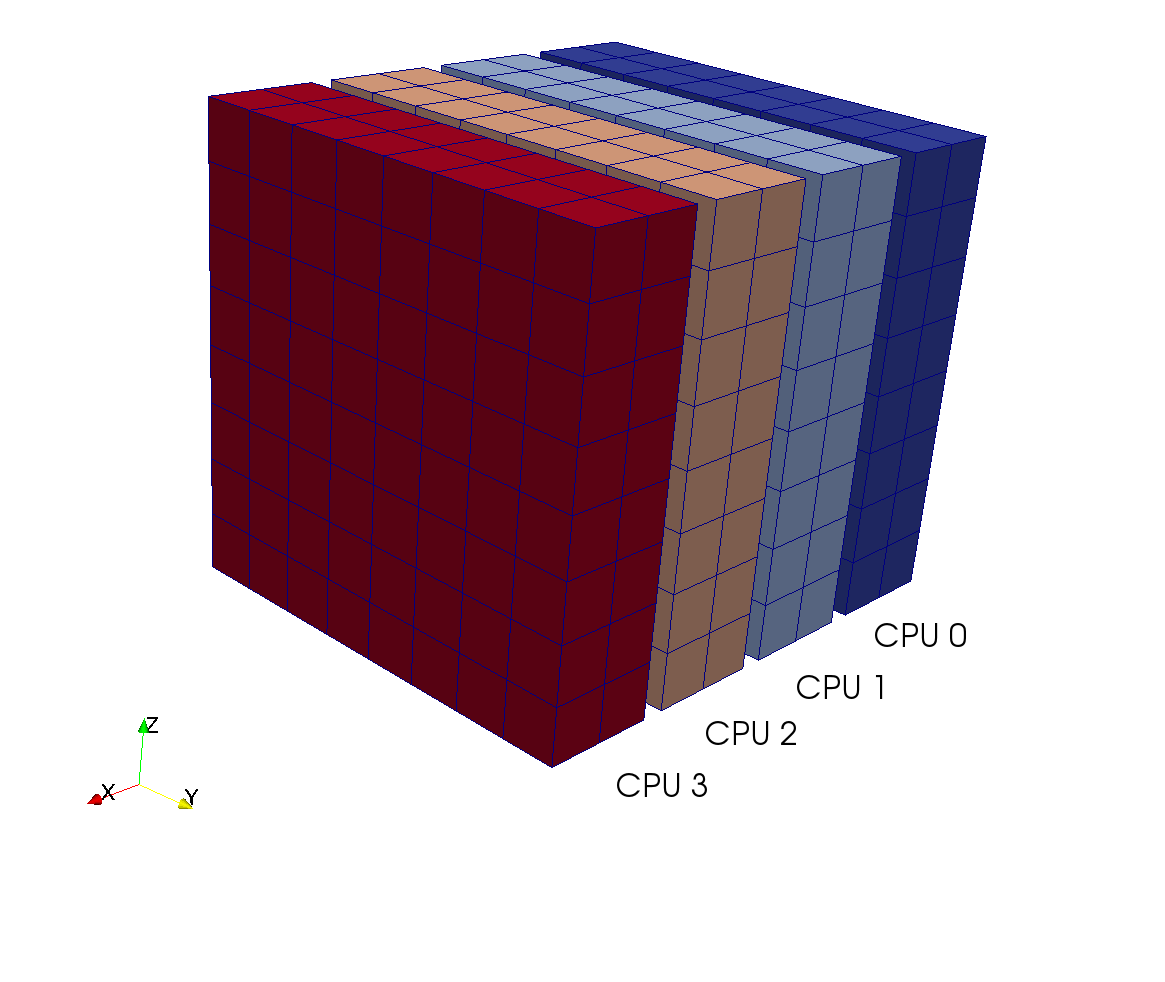}
  \label{slabsubfig0}
  }
\subfigure[Wavenumber mesh after real transform]{
  \includegraphics[scale=0.15]{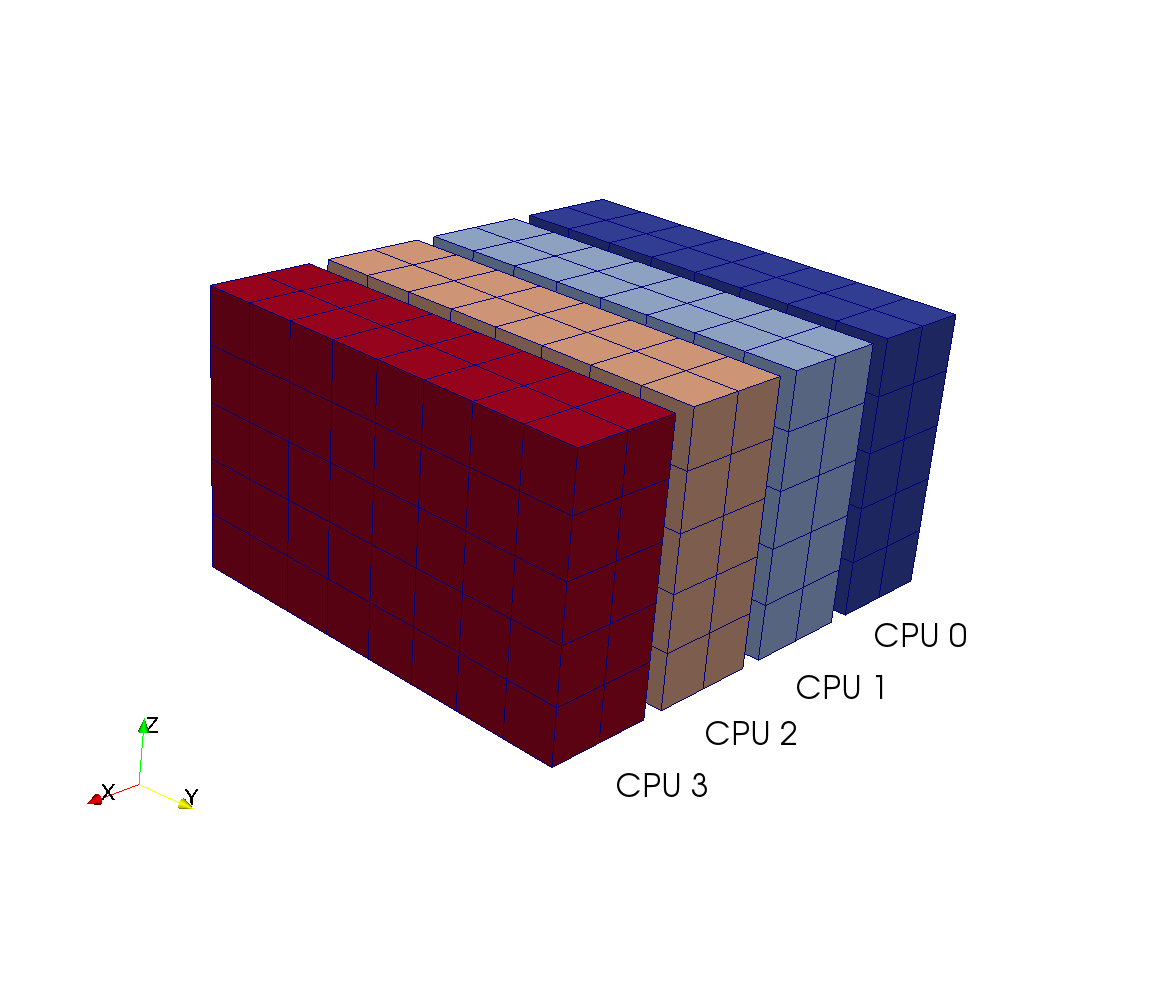}
  \label{slabsubfig1}
}
\subfigure[Final wavenumber mesh.]{
  \includegraphics[scale=0.15]{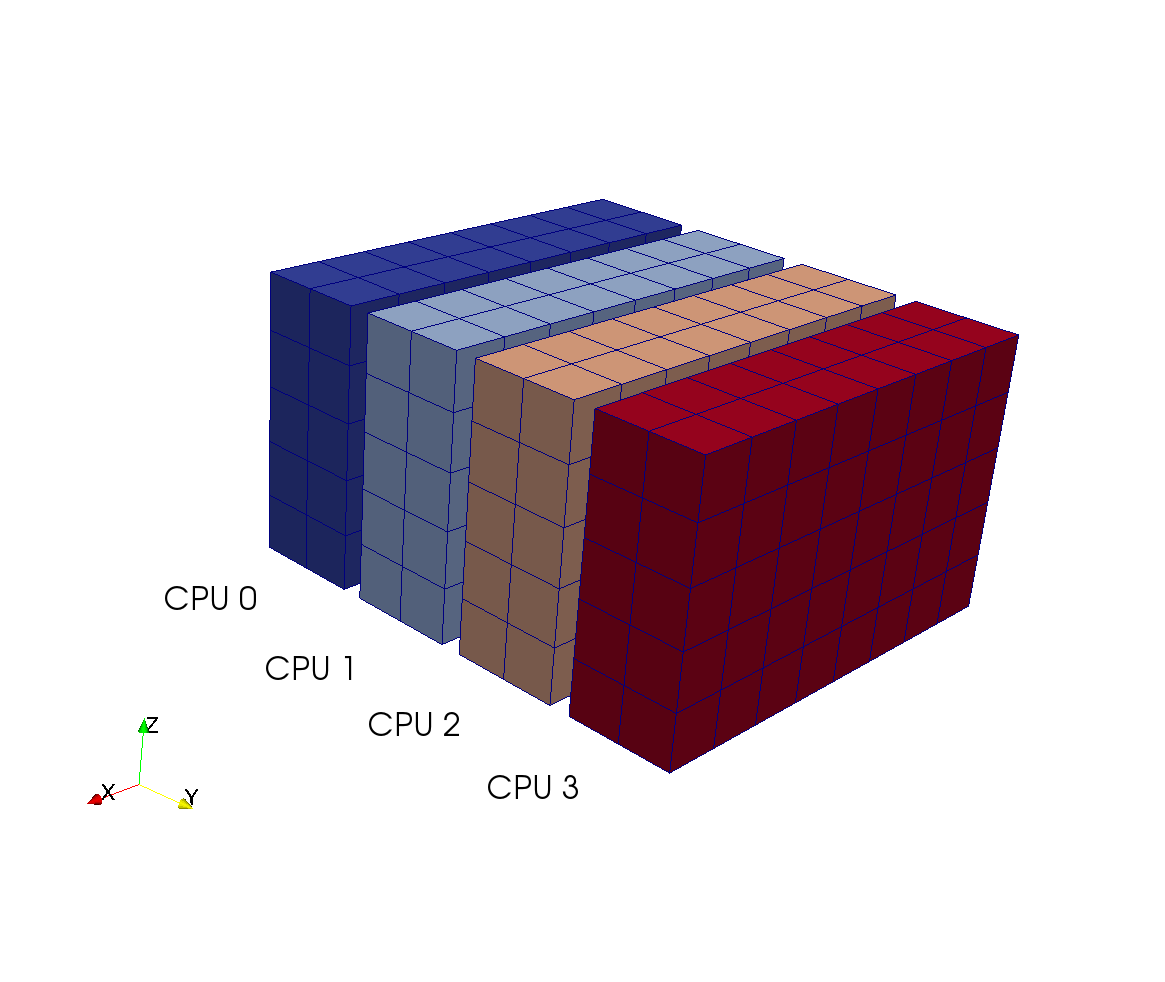}
  \label{slabsubfig2}
  }
\caption{Slab decomposition of physical mesh \subref{slabsubfig0}, intermediate 
wavenumber mesh \subref{slabsubfig1}, and final wavenumber mesh 
\subref{slabsubfig2}. In \inpyth{fftn_mpi} from Figure \ref{fig:fftn_slab} the 
function \inpyth{rfft2} 
transforms in $y$ and $z$-directions and moves the real data in 
\subref{slabsubfig0} to the complex datastructure in \subref{slabsubfig1}. The 
complex data is then transposed and communicated to reach the final structure 
seen in \subref{slabsubfig2}. }
\label{fig:Slabdecomp}
\end{figure*}

\subsection{2D Pencil decomposition}
\label{pencil2D}

For massively parallel simulations the slab decomposition falls short since the number of CPUs allowed is limited by $N$. Large-scale simulations using up to $N^2$ CPUs commonly employ the 2D pencil decomposition, first suggested by Ding, Ferraro and Gennery in 1995 \cite{Ding95}. Publically available implementations of the 3D parallel FFT that makes use of the pencil decomposition are the Parallel FFT Subroutine Library by Plimpton \cite{PlimptonFFT}, the P3DFFT library by Pekurovsky \cite{p3dfft, pekurovsky2012}, the 2DECOMP\&FFT library by Li and Laizet \cite{Li2010} and PFFT by Pippig \cite{Pi13}.

\begin{figure*}[th!]
\centering
\subfigure[Physical mesh]{
  \includegraphics[scale=0.15]{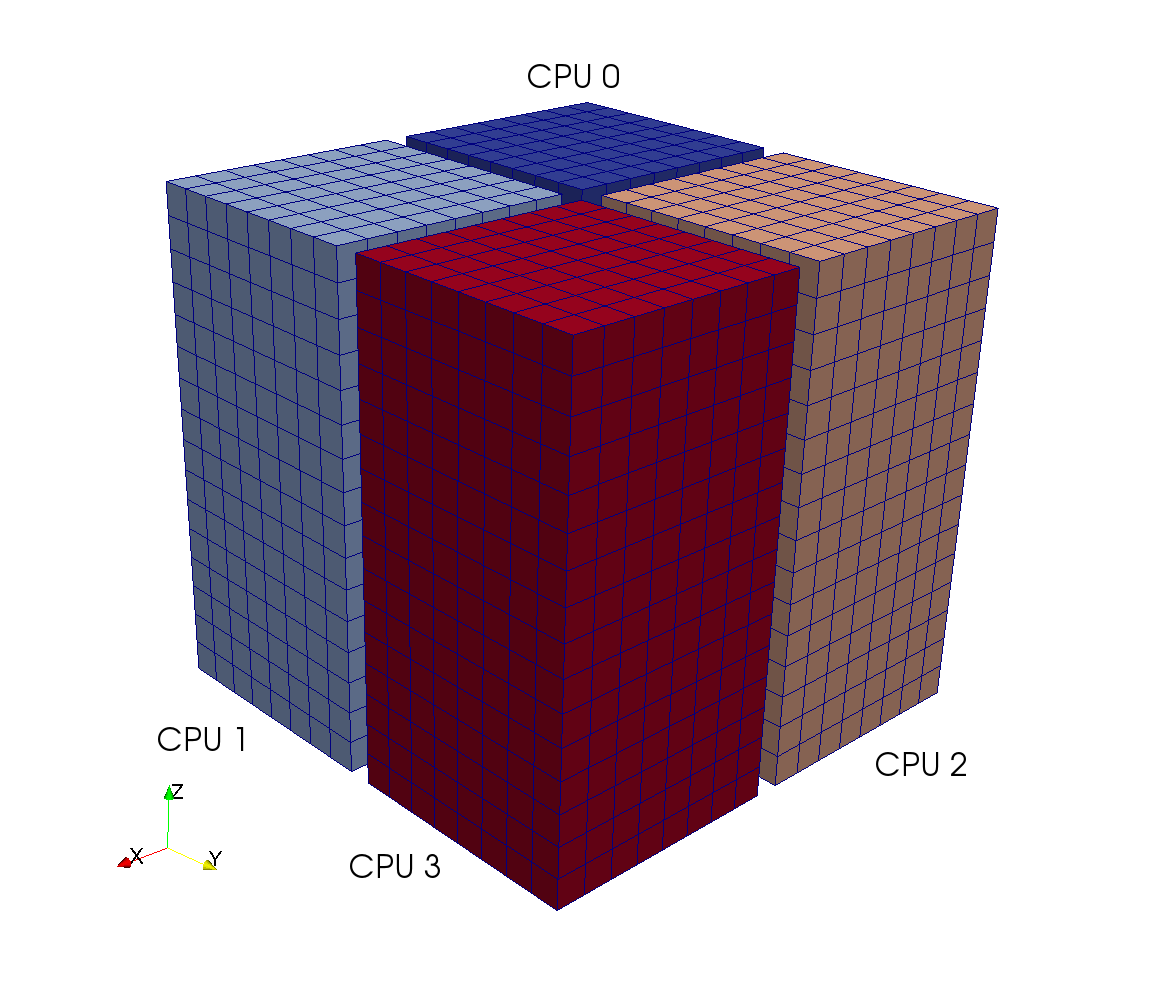}
  \label{subfig0}
  }
\subfigure[Wavenumber mesh after real transform]{
  \includegraphics[scale=0.15]{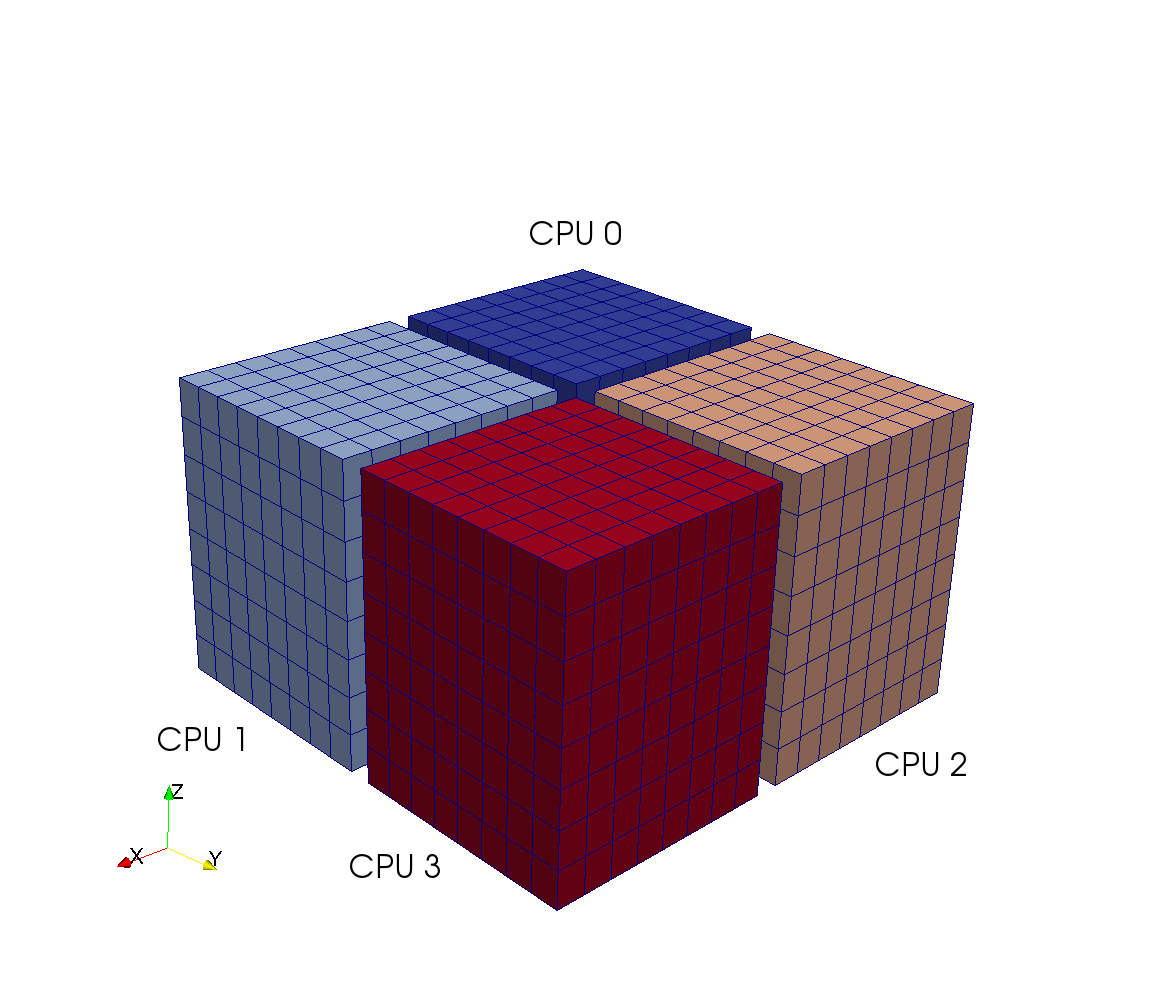}
  \label{subfig1}
  }
\subfigure[Intermediate wavenumber mesh]{
  \includegraphics[scale=0.15]{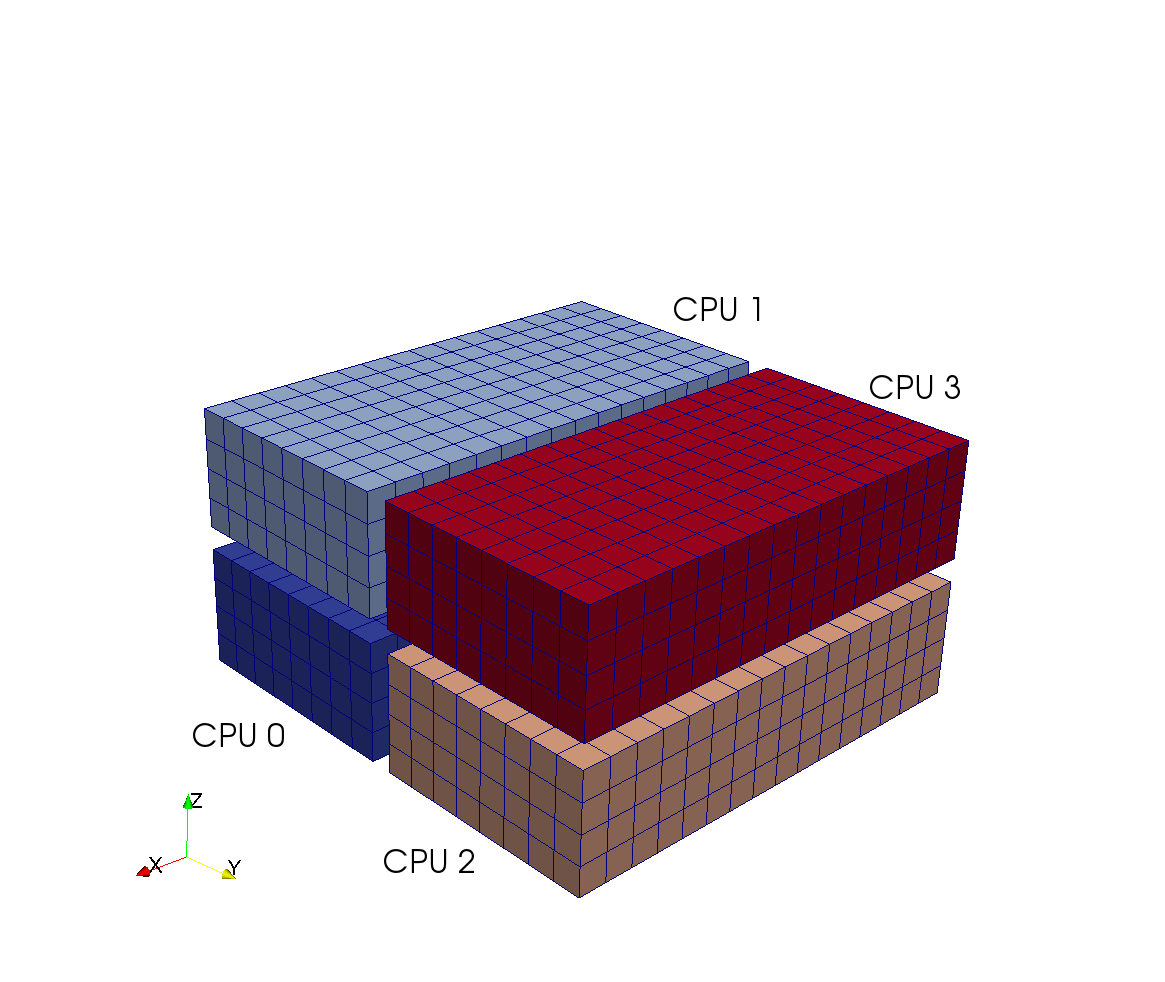}
  \label{subfig2}
  }
\subfigure[Final wavenumber mesh.]{
  \includegraphics[scale=0.15]{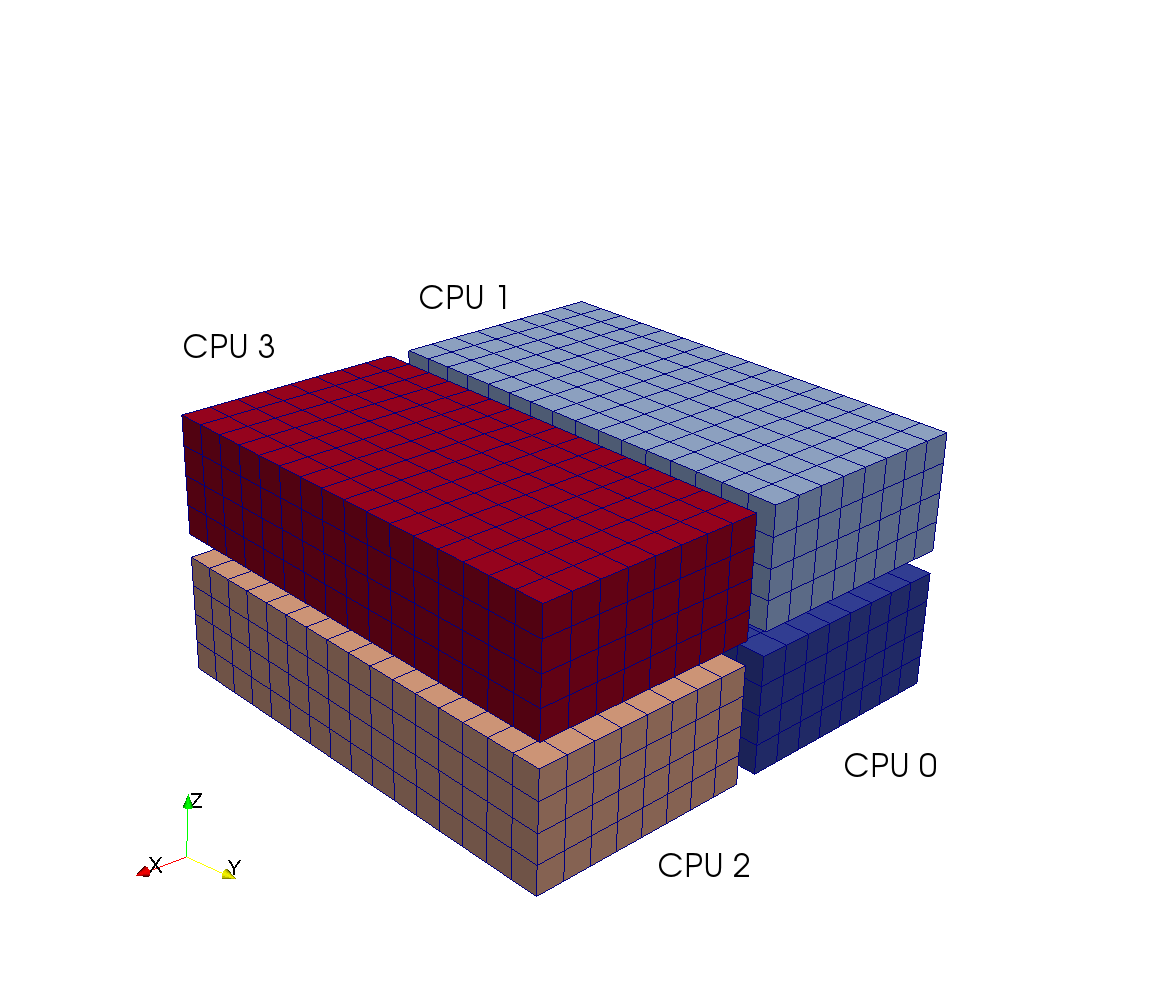}
  \label{subfig3}
  }
\caption{2D pencil decomposition of physical mesh \subref{subfig0} and the three wavenumber meshes \subref{subfig1}, \subref{subfig2}, \subref{subfig3}. The decomposition shown uses 4 CPUs, two in each direction normal to the direction of the current one-dimensional FFT. The FFT in the $z$-direction transforms the real data in \subref{subfig0} to the complex data in \subref{subfig1}. The data is then transformed and communicated to the composition seen in \subref{subfig2}. Here the FFT in $x$-direction is carried out before the data is transformed again and communicated to \subref{subfig3}, where the final FFT is performed.}
\label{fig:Pencildecomp}
\end{figure*}

The 2D pencil decomposition strategy is illustrated in 
Figure~\ref{fig:Pencildecomp} for a physical computational box of size $16^3$, 
using 4 CPUs. The datastructures are split in the plane normal to the direction 
in which the FFT is performed. That is, for the physical mesh in 
Figure~\ref{fig:Pencildecomp} (a) the $x-y$ plane is split up in a $2\times2$ 
processor mesh. Each CPU may now perform 64 ($= 8 \times 8$) real FFTs of 
length 16 in the $z$-direction on its own $8 \times 8 \times 16$ physical mesh. 
Afterwords, the complex data are laid out as shown in 
Figure~\ref{fig:Pencildecomp} (b). The second transform takes place in the 
$x$-direction, and for this to happen, data must be exchanged between 
processors 0 and 1 as well as 2 and 3. The datastructures must also be 
transformed to the shape seen in Figure~\ref{fig:Pencildecomp} (c). Each CPU 
may now perform the 32 ($4 \times 8$) individual 1D FFTs in the $x$-direction. 
The same procedure is followed to end up with datastructures aligned in the 
final $y$-direction, see Figure~\ref{fig:Pencildecomp} (d). However, this time 
processor 0 communicates with processor 2 and likewise 1 with 3. Evidently, 
each processor belongs to two different groups of communicators. One for 
communication in the $x-z$ plane (Figure~\ref{fig:Pencildecomp} (b) to (c)) and 
one for communication in the $x-y$ plane (Figure~\ref{fig:Pencildecomp} (c) to 
(d)). The MPI communicator groups and the distributed mesh are created in 
Python as shown in Figure \ref{fig:pencil_init}

%\noindent
%\begin{minipage}[l]{\textwidth}
\begin{figure}
\begin{python}
P1 = 2  # CPUs in 1st direction (assigned)
P2 = num_processes / P1 # CPUs in 2nd dir
N1 = N/P1
N2 = N/P2

# Create two communicator groups
commxz = comm.Split(rank/P1)
commxy = comm.Split(rank%P1)

# Get local ranks
xzrank = commxz.Get_rank()
xyrank = commxy.Get_rank()

# Create the decomposed physical mesh
x1 = slice(xzrank * N1, (xzrank+1) * N1)
x2 = slice(xyrank * N2, (xyrank+1) * N2)
X = mgrid[x1, x2, :N].astype(float)*L/N

# Create the decomposed wavenumber mesh
k2 = slice(xyrank*N2, (xyrank+1)*N2)
k1 = slice(xzrank*N1/2, (xzrank+1)*N1/2)
K = array(meshgrid(kx[k2], kx, kx[k1], 
          indexing='ij'), dtype=int)
\end{python}
\caption{Creation of MPI communicator groups and physical and wavenumber meshes 
for pencil decomposition.}
\label{fig:pencil_init}
\end{figure}	

%\end{minipage}

With reference to Figure~\ref{fig:Pencildecomp}, the two communicator groups for CPU with global rank 0 is
\inpyth{commxz} = [0, 1] and \inpyth{commxy} = [0, 2]. The decomposition contains two parameters, \inpyth{P1} and \inpyth{P2}, that are used to split up the first two indices of the physical mesh.
For the case shown \inpyth{P1} may be either 1 or 2, in which case the second parameter \inpyth{P2}
is 4 or 2 to arrive at a total of \inpyth{num_processes = P1*P2}.

The entire 7 lines of code required for the Python implementation of the 3D 
parallel FFT 
with the 2D pencil decomposition appears in Figure \ref{fig:fftn_pencil}. The 
work arrays \inpyth{Uc_hat_y, Uc_hat_x, Uc_hat_z} are laid out as seen in 
Figure~\ref{fig:Pencildecomp} (d), (c) and (b) respectively. The array
\inpyth{Uc_hat_xr} is a copy of \inpyth{Uc_hat_x} used only for communication. 
Note that all MPI communications take place with the data aligned in the 
$x$-direction, as shown in Figure \ref{fig:Pencildecomp} (c), because this 
requires no preparation for the \inpyth{Alltoall} call. The first
index is automatically split up and distributed.  In the call to 
\inpyth{Alltoall} the first dimension of \inpyth{Uc_hat_x}, i.e. $N$, is 
automatically broadcasted such that for \inpyth{commxz} \inpyth{Uc_hat_x}
gets the shape \inpyth{(P1, N1, N2, N1/2)} and \inpyth{commxy} \inpyth{(P2, N2, N2, N1/2)}.

Note that the Nyquist frequency has been neglected for the pencil 
decomposition. This is not uncommon in turbulence simulations, because the 
Nyquist frequency ($k=N/2+1$) ``is a coefficient for a Fourier mode
that is not carried in the Fourier representation of the solution'' 
\cite{Lee2013}.  Another more practical reason is that the number of 
independent wavenumbers in the z-direction is odd and thus not possible to 
share evenly between an even number of processors. There are, however, several 
different solutions to this problem, besides neglecting the Nyquist frequency. 
One solution is to let one processor have responsibility for one
more wavenumber than the others, which leads to slightly poor load balancing 
(increasingly better for large number of processors). A better choice then is 
to place the real numbers of the Nyquist frequency in the complex part of the 
lowest wavenumber (which is also real) before transforming and communicating. 
This is done by Pekurovsky et~al. \cite{pekurovsky2012} in their P3DFFT code. 
The approach is quite easily implemented in Python, taking no more than 10 
extra lines of code for both the forward and the backward
transforms. However, there is additional communication and the computational 
cost is slightly higher. For this reason, we have here simply chosen to neglect 
the Nyquist frequency.

\begin{figure*}
\begin{python}
def fftn_mpi(u, fu):
    """fft in three directions using MPI and the pencil decomposition"""
    Uc_hat_z[:] = rfft(u, axis=2)

    # Transform to x direction neglecting k=N/2 (Nyquist)
    Uc_hat_x[:] = rollaxis(Uc_hat_z[:,:,:-1].reshape((N1, N2, P1, N1)), 2). \
                                             reshape(Uc_hat_x.shape)

    # Communicate and do fft in x-direction
    commxz.Alltoall([Uc_hat_x, mpitype], [Uc_hat_xr, mpitype])
    Uc_hat_x[:] = fft(Uc_hat_xr, axis=0)

    # Communicate and transform to z-direction
    commxy.Alltoall([Uc_hat_x, mpitype], [Uc_hat_xr, mpitype])
    Uc_hat_y[:] = rollaxis(Uc_hat_xr.reshape((P2, N2, N2, N1)), 1). \
                                     reshape(Uc_hat_y.shape)

    fu[:] = fft(Uc_hat_y, axis=1)
    return fu
\end{python}
\caption{Three dimensional forward \inpyth{fftn_mpi}  FFT implemented for the 
pencil decomposition.}
\label{fig:fftn_pencil}
\end{figure*}

\subsection{Optimization}
\label{sec:optimization}
The scientific Python implementation discussed thus far is using no more than 
NumPy and MPI for Python, and possibly pyFFTW for faster FFTs. The FFTs 
represent the major computational cost of the solver, but there is also 
significant cost in all the elementwise operations on the structured NumPy 
arrays required to build the right hand side of the Navier-Stokes equations (
see the \inpyth{computeRHS} function). The elementwise operations, like 
multiply 
and divide, are handled by NumPy's universal functions (ufuncs), that are fast 
because they are implemented in C. However, some of the ufuncs may be 
inefficient due to temporary arrays that need to be created, especially for 
multiple-operator expressions. The most glaring example for the current solver 
is the cross product (used in computing convection, see Eq. 
(\ref{eq:curl_convection})), which is computed most straightforward though 
NumPy as
\begin{python}
def cross(a, b, c):
    """c = a x b"""
    #Alternative numpy cross product:
    #c = numpy.cross(a, b, axis=0)
    c[0] = a[1]*b[2]-a[2]*b[1]
    c[1] = a[2]*b[0]-a[0]*b[2]
    c[2] = a[0]*b[1]-a[1]*b[0]
    return c
\end{python}
Note that we have commented out the use of NumPy's \inpyth{cross} function, 
which is implemented under the hood using ufuncs and runs at approximately the 
same computational speed as the alternative shown. The term 
\inpyth{a[1]*b[2]-a[2]*b[1]} requires two elementwise multiplications and one 
elementwise subtraction. This is implemented by NumPy using three separate 
ufuncs (two \inpyth{multiply}, one \inpyth{subtract}), each with temporary 
arrays allocated. For the three items in the \inpyth{c} vector, this means that 
9 ufuncs are called and the loop over all $N^3$ indices in the mesh is 
performed 9 times, which, needless to say, is very inefficient even though it 
is vectorized.

For straightforward comparison, the Python solver with slab decomposition has been used as prototype for a
\href{https://github.com/mikaem/spectralDNS/blob/master/cpp/spectralDNS.cpp}{pure
 C++ 
implementation}\footnote{\texttt{https://github.com/mikaem/spectralDNS/blob/master
 /cpp/spectralDNS.cpp}}. The
only significant difference between the solvers is that for C++, the 3D FFT 
computations are using the MPI communication routines that are provided 
directly through the FFTW library. Otherwise, everything is exactly the same. 
For a first direct comparison, though, we are neglecting the MPI implementation 
and running the Python and C++ codes in serial for the Taylor Green test case 
(see Section \ref{sec:verification}) for three meshes of size $(32^3, 64^3, 
128^3)$. On a
MacBook Pro with 2.8 GHz Intel Core i7 we find that the C++ code runs at (0.0092, 0.083, 0.80) seconds per time step for the three mesh sizes, whereas the pure Python solver runs at (0.013, 0.12, 1.3) seconds.\footnote{Best of 10 runs} As such we may conclude that without MPI the C++ code is approximately 30 \% faster. Considering the issue recently discussed with the cross product, we are all in all quite happy with the performance of the Python solver.

Fortunately, many different strategies are known for speeding up pure 
NumPy code, e.g.,  Numba \cite{numba}, Numexpr \cite{numexpr}, Weave 
\cite{weave}
and Cython \cite{cython}. We have obtained best results for Numba 
and Cython. The {Numba} module uses just in time 
compilation of Python functions chosen for optimization. The advantage of using 
Numba is that no compilation or installation is required by the user 
who never has to leave the pure Python environment. The downside is that 
{Numba} is difficult to install on non-standard platforms like the Blue 
Gene/P. Nevertheless, a {Numba} optimized cross product routine may be 
implemented in an external module and code will be generated and 
compiled when this module is imported.  For efficient speed-up on our 
structured 
mesh the code must be rewritten using explicit for loops as seen in Figure 
\ref{fig:numba}. The {Numba} cross product may now be used directly in 
place of the pure Python version with no further modifications.
\begin{figure}
\begin{python}
from numba import jit, float64 as float

@jit(float[:,:,:,:](float[:,:,:,:], 
     float[:,:,:,:], float[:,:,:,:]))
def cross(a, b, c):
    for i in xrange(a.shape[1]):
        for j in xrange(a.shape[2]):
            for k in xrange(a.shape[3]):
                a0 = a[0,i,j,k]
                a1 = a[1,i,j,k]
                a2 = a[2,i,j,k]
                b0 = b[0,i,j,k]
                b1 = b[1,i,j,k]
                b2 = b[2,i,j,k]
                c[0,i,j,k] = a1*b2 - a2*b1
                c[1,i,j,k] = a2*b0 - a0*b2
                c[2,i,j,k] = a0*b1 - a1*b0
    return c
\end{python}
\caption{Numba implementation of the cross product in double 
precision.}\label{fig:numba}
\end{figure}

Cython has gained much momentum in recent years and is usually the obvious 
choice for high performance computing. Using Cython, the cross product should 
be implemented like for {Numba}, through explicit for loops which are 
translated to efficient C code to be highly optimized by the compiler. As shown 
in Figure \ref{fig:cython} an external module may be created and compiled using 
Python's standard installation utilities 
(\href{http://docs.cython.org/src/reference/compilation.html}{distutils}\footnote{\texttt{http://docs.cython.org/src/reference/compilation.html}}).
\vskip 1ex
\begin{figure}
\begin{python}
#cython: boundscheck=False
#cython: wraparound=False
cimport numpy as np

ctypedef np.float64_t real_t

def cross(np.ndarray[real_t, ndim=4] a,
          np.ndarray[real_t, ndim=4] b,
          np.ndarray[real_t, ndim=4] c):
    cdef unsigned int i, j, k
    cdef real_t a0, a1, a2, b0, b1, b2
    for i in xrange(a.shape[1]):
        for j in xrange(a.shape[2]):
            for k in xrange(a.shape[3]):
                a0 = a[0,i,j,k]
                a1 = a[1,i,j,k]
                a2 = a[2,i,j,k]
                b0 = b[0,i,j,k]
                b1 = b[1,i,j,k]
                b2 = b[2,i,j,k]
                c[0,i,j,k] = a1*b2 - a2*b1
                c[1,i,j,k] = a2*b0 - a0*b2
                c[2,i,j,k] = a0*b1 - a1*b0

\end{python}
\caption{Cython implementation of the cross 
product in double precision.}\label{fig:cython}
\end{figure}
Cython generated modules may be imported into Python and the \inpyth{cross} 
function from Cython may be used directly in place of the pure Python version. 
The Cython code for the cross product runs at about the same speed as 
Numba and approximately 3 times faster than pure NumPy for a mesh of 
size $128^3$. When Cython is used to compute all elementwise operations that 
generate the right hand side of the Navier-Stokes equations, the Taylor Green 
test case runs at (0.0090, 0.080, 0.82) seconds per time step.\footnote{Best of 
10 runs} In other words, with a few quick Cython wrappers we have a Python 
solver that is comparable in speed to its low-level counterpart in C++. And 
this solver is scripted and operated in a comfortable interactive environment 
where post processing and plotting may be performed from a commandline shell.

\section{Parallel performance}
Our scientific Python DNS solver has been tested on the Shaheen computer at the 
KAUST Supercomputing Laboratory. The primary computational resource of Shaheen 
consists of 16 racks of Blue Gene/P. Each rack contains 1024 quad-core, 32-bit, 
850 MHz PowerPC compute nodes, for a total of 65,536 compute cores. Each node 
is equipped with 4GB of system memory, providing a total 64TB of distributed 
memory across the resource. Blue Gene nodes are interconnected by a 
three-dimensional point-to-point ``torus'' network used for general-purpose 
message-passing and multicast communication. 

%By using many small, low-power, densely packaged chips, Blue Gene/P exceeded the power efficiency of other supercomputers of its generation, and at 371 MFLOPS/W Blue Gene/P installations ranked at or near the top of the Green500 lists in 2007-2008 \cite{top500green}.  Per November 2009, the TOP500 list contained 15 Blue Gene/P installations of 2-racks (2048 nodes, 8192 processor cores, 23.86 TFLOPS Linpack) and larger \cite{top500}.

\subsection{Dynamic loading of Python}
Running a Python solver on a supercomputer presents a few challenges. Most 
importantly, the dynamic loading time of Python, i.e., the time to simply get 
started, may become unreasonably high. To load Python with NumPy on the 
complete 16 racks of Shaheen takes approximately 45 minutes. The main reason is 
that large-scale supercomputers, such as Shaheen, make use of parallel file 
systems designed to support large distributed loads, where each CPU is writing 
or reading data independently. With dynamic loading, however, each process 
attempts to access the same files simultaneously and bottlenecks appear.

A few solutions to the loading problem are known. The \inpyth{MPI_Import} 
Python module of Langton \cite{mpi_import} is designed to overload the built in 
import mechanism of Python such that only one rank (the zero rank) searches the 
filesystem for a module and then subsequently broadcasts the results to the 
remaining ranks. A drawback is the Python implementation where several modules 
(e.g., \inpyth{sys, imp, __builtin__,types, mpi4py}) need to be imported 
through the regular Python import mechanism. A second approach of Langton is to 
rewrite the regular finders and loaders modules such that one rank does all the 
initial work, caching the modules it finds.

A better solution is provided through the Scalable Python implementation 
\cite{scalablepython, Enkovaara201117}, used, e.g., by GPAW \cite{gpaw05}. Here 
CPython is modified slightly such that during import operations only a single 
process performs the actual I/O, and MPI is used for broadcasting the data to 
other MPI ranks. The solution provided by Scalable Python is used in this work 
and as such dynamic loading times are kept very low (approximately 30 seconds 
for a full rack).

\subsection{Verification}
\label{sec:verification}
To verify the implementation of the scientific Python DNS solver we run the 
Taylor Green test case for a Reynolds number of 1600. This is a well known test 
case used, e.g., by the annual Workshop on High-Order CFD Methods, which 
provides reference 
data\footnote{https://www.grc.nasa.gov/wp-content/uploads/sites/22/ 
C3.3$\_$datafiles.zip} for the temporal evolution of kinetic energy, energy 
dissipation and vorticity 
integrated over the computational domain. The Taylor Green test case at Re=1600 
starts out as laminar with only one single length scale, and then through a 
series of complex flow patterns a fully developed, decaying, turbulent flow is 
obtained. The Taylor Green vortex is initialized as
\begin{python}
nu = 1./1600
dt = 0.001
U[0] = sin(X[0])*cos(X[1])*cos(X[2])
U[1] =-cos(X[0])*sin(X[1])*cos(X[2])
U[2] = 0
for i in range(3):
    U_hat[i] = fftn_mpi(U[i], U_hat[i])
\end{python}
where the same code is used for both the slab and the pencil decomposition. The Reynolds number is fixed at 1600 by setting the viscosity coefficient to 1/1600 and the time step is fixed at 0.001. We use a real mesh of size $N=512$, since that matches the reference data. A resolution of $512^3$ is reported by by Bratchet \cite{brachet1991direct} to represent a fully resolved DNS of this flow.

The non-optimized code is run on Shaheen using the pencil decomposition and 2048 cores. The entire simulation is set for 20 seconds real time, during which the flow undergoes transition, decays and finally fades out. The simulation takes 12 hours alltogether ($12\cdot2048$ CPU hours) and neither the kinetic energy nor the vorticity stray more than 0.01 \% from the reference solutions, even after 20,000 time steps.

\subsection{Parallel scaling}
In this section we will investigate parallel scaling and performance of the 
solver described in previous sections. Optimization of the 
scientific Python solver will be based on Cython and not Numba, because we only 
managed to install the former on the Blue Gene nodes. We compare our Python 
based solvers to the C++ solver described briefly in Sec \ref{sec:optimization} 
and note that the C++ solver is using the same compiled FFTW library as the 
Python code. However, it is also important to note that the Python code only 
makes use of the serial versions of the FFTW routines and performs the MPI 
communications itself through MPI fo Python, whereas the C++ code uses the 
built in MPI communication routines that come with the FFTW library.

The only MPI communication required by the current solver can be found in the 3D
FFT routines, or when global results, like the total kinetic energy, are
computed. Scaling will be efficient as long as the cost for MPI
communication is smaller than all the other elementwise operations required
to set up the right hand side of the explicit ODE given by Eq.
(\ref{eq:NSfinal}). This is not a trivial task, considering that a 
pseudo-spectral solver demands that every process sends and receives data
from all the other processes - at least for the slab decomposition.

It is customary for pseudo-spectral solvers to assume that the main computational cost is due to Fourier transforms, and as such the CPU time should ideally scale like

\begin{equation}
 \frac{N^3 \log_2 N}{M},
\end{equation}
where $M$ is the number of CPUs. In other words, strong scaling implies, as 
usual, CPU time per time step proportional to $1/M$, whereas weak scaling, 
where $N^3/M$ is kept constant, should ideally show CPU times proportional to 
$\log_2 N$. A more thorough analysis of scalability factors for a 3D FFT using 
the pencil domain decomposition has been given by Ayala and Wang 
\cite{ayala2013}. Note that all results are computed using single precision 
arithmetics, which is customary for scaling tests of similar DNS solvers. 

\subsubsection{Weak scaling}
Since MPI communications are primarily required by the FFTs, we first run a 
weak scaling test of the pure forward and inverse FFTs before proceeding to the 
complete solver. To this end we keep the ratio of mesh size to number of 
processors ($N^3/M$) constant, which ideally should lead to CPU times 
proportional to $\log_2 N$. Both slab and pencil decompositions are tested 
using four meshes of size $N^3$, where $N=(2^7, 2^8, 2^9, 2^{10}) =(128, 256, 
512, 1024)$. For these meshes the slab decomposition uses 
$M=(2, 16, 128, 1024)$ number of CPUs and the pencil uses twice as many $M=(4, 
32, 256, 2048)$ with $P1=(2, 4, 16, 32)$.
Note that for $N=1024$ the maximum number of CPUs is reached for the slab decomposition, whereas the pencil
decomposition may still use many more. The performance of the FFTs is compared 
to a C++ code using the same FFTW library with plans corresponding to the 
Python solver. A simplified setup that also shows how the code is executed is 
shown in Figure \ref{fig:C++FFTW}. The FFTW library is using a slab mesh 
decomposition.

\begin{figure*}[ht!]
\begin{lstlisting}
  typedef float precision;
  int N = pow(static_cast<int>(2), 9);
  ptrdiff_t alloc_local, local_n0, local_0_start, local_n1, local_1_start
  fftwf_plan rfftn, irfftn;

  alloc_local = fftw_mpi_local_size_3d_transposed(N, N, N/2+1,
                         MPI::COMM_WORLD, &local_n0, &local_0_start,
                         &local_n1, &local_1_start);

  vector<precision> U(2*alloc_local);
  vector<complex<precision> > U_hat(alloc_local);

  rfftn = fftwf_mpi_plan_dft_r2c_3d(N, N, N, U.data(),
                   reinterpret_cast<fftwf_complex*>(U_hat.data()),
                   MPI::COMM_WORLD, FFTW_MPI_TRANSPOSED_OUT);

  irfftn = fftwf_mpi_plan_dft_c2r_3d(N, N, N,
                   reinterpret_cast<fftwf_complex*>(U_hat.data()),
                   U.data(), MPI::COMM_WORLD, FFTW_MPI_TRANSPOSED_IN);

  // Execute forward and inverse transforms
  fftwf_mpi_execute_dft_r2c( rfftn, U.data(),
                reinterpret_cast<fftwf_complex*>(U_hat.data()));
  fftwf_mpi_execute_dft_c2r(irfftn,
                reinterpret_cast<fftwf_complex*>(U_hat.data()), U.data());

  // Scale the inverse FFT
  for (int k=0; k<U.size(); k++)
     U[k] /= tot;

\end{lstlisting}
\caption{C++ setup for the parallel FFT using FFTW.}
\label{fig:C++FFTW}
\end{figure*}

%The slab decomposition is seen to be more efficient than pencil, which is generally in agreement with previous studies (see, e.g., \cite{pekurovsky2012}). We see also that the C++ version is faster than the corresponding Python routine (slab) by approximately 10 \% for the lowest number of processors (64). However, the strong scaling of the Python 4-liner slab (see \inpyth{fftn_mpi}) is better and the Python version is even faster by the time we reach the highest number of allowed processors for this case (512). The pencil transform achieves good, but not perfect, speedup all the way up to 4096 processors. The results for the pencil transforms depend on the parameter P1, set to determine how the two first indices of the computational mesh are distributed. We have used $P1 = (8, 16, 16, 32, 32, 64, 64)$ and not optimized further.

\begin{figure}[ht!]
\centering
\includegraphics[scale=0.48]{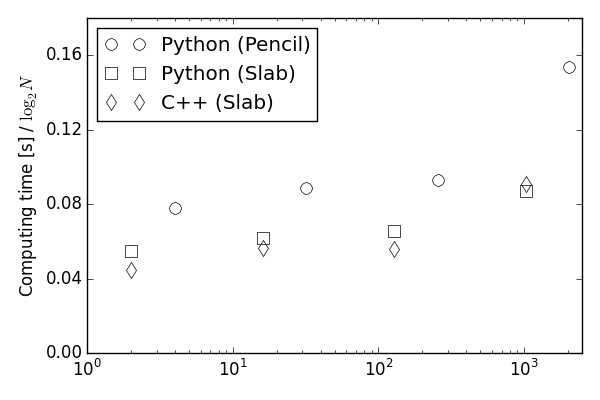}
\caption{Weak scaling of forward and inverse FFTs. The scientific Python 
results are not optimized and C++ is using slab 
decomposition with MPI communication performed through the FFTW library. The 
pencil timings are scaled by a factor 0.5 since the pencil results are 
collected using twice as many cores as slab.}
\label{fig:weak_FFT_scaling}
\end{figure}

The results are collected as the fastest of 10 forward and inverse FFTs (only 
execution time, not the setup), and the timings
are scaled by $\log_2N$ and plotted in Figure \ref{fig:weak_FFT_scaling}. 
Perfect scaling would thus imply a constant compute time as the number of cores 
is increased. From Figure \ref{fig:weak_FFT_scaling} we observe that both slab 
and pencil versions of the Python 3D FFTs show good scaling,
and only start to trail off significantly for the largest mesh size. The C++ 
routine can be seen to be slightly faster than Python (slab) for the 
smallest number of processors (64), whereas the Python routine is fastest for 
the largest number of cores (1024). In other words, the Python 4-liners (see 
Figure \ref{fig:fftn_slab}) that embody the complete execution of the 3D FFT 
with MPI scales slightly better than the 3D FFT routines with MPI provided by 
FFTW.  

From Figure \ref{fig:weak_FFT_scaling} we also observe that the slab 
decomposition is more efficient than pencil, which is generally in agreement 
with previous studies (see, e.g., \cite{pekurovsky2012}). Note that the pencil 
computations are using twice as many cores for the same mesh as the slab 
computations. As such the pencil results in Figure \ref{fig:weak_FFT_scaling} 
have been scaled by a factor 0.5 to allow for a direct comparison. 

\begin{figure}[ht!]
\subfigure[Scientific Python solver]{
\includegraphics[scale=0.44]{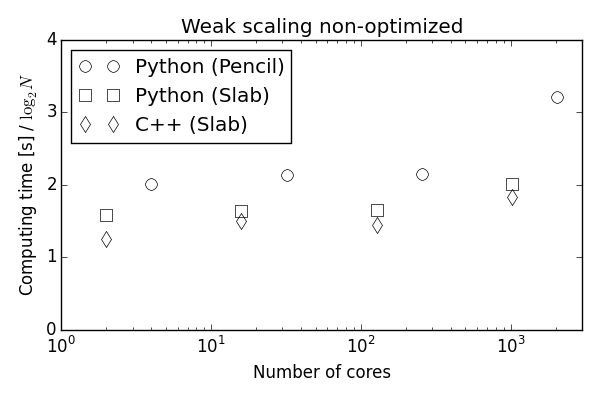}
\label{fig:weak_scaling_shaheen}
}
\subfigure[Python solver optimized with Cython]{
\includegraphics[scale=0.44]{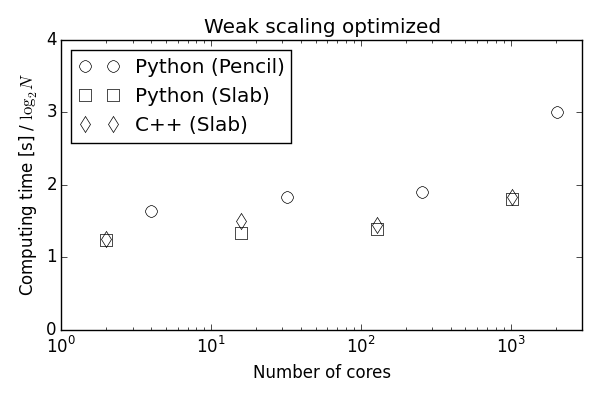}
\label{fig:weak_scaling_shaheen_opt}
}
\caption{Weak scaling of DNS solver. The slab decompositions are using $4 \cdot 64^3$ nodes per core, whereas the pencil decomposition uses $2 \cdot 64^3 $. The C++ solver uses slab decomposition with MPI communication performed through the FFTW library. The C++ results are the same in \subref{fig:weak_scaling_shaheen} and \subref{fig:weak_scaling_shaheen_opt}.}
\label{fig:Weak_scaling}
\end{figure}

Turning now to the complete solver, we run the Taylor Green test case for 100 
time steps on the same meshes that were used for the weak scaling of the FFTs. 
Figure \ref{fig:Weak_scaling} shows the computing times for one such time step 
(the fastest) for both the scientific Python solver 
\ref{fig:weak_scaling_shaheen} 
and the version optimized with Cython \ref{fig:weak_scaling_shaheen_opt}.
The figures clearly show how the Cython optimization leads to a solver that 
performs as well as C++ for the entire range of processors. Both the slab and 
pencil optimized implementations are seen to be approximately $20 \% $ faster 
than their non-optimized Python counterparts, but there is generally no 
difference in scaling performance. Like for the pure FFT test, the scaling 
performance of the Python solver is actually slightly better  than C++, and 
this leads to the optimized Python solver actually being the fastest for 1024 
processors.

\subsubsection{Strong scaling}
\label{sec:strong_scaling}

Strong scaling over a wide range of processors is usually more challenging
to prove than weak, in our case partly because the efficiency of the FFT 
routines depends on the length of the datastructures and how well they fit into
memory. Nevertheless, we have performed strong scaling tests of the FFTs and the complete solvers using a computational box of size $512^3$. 

\begin{figure}[ht!]
	\centering
	\includegraphics[scale=0.44]{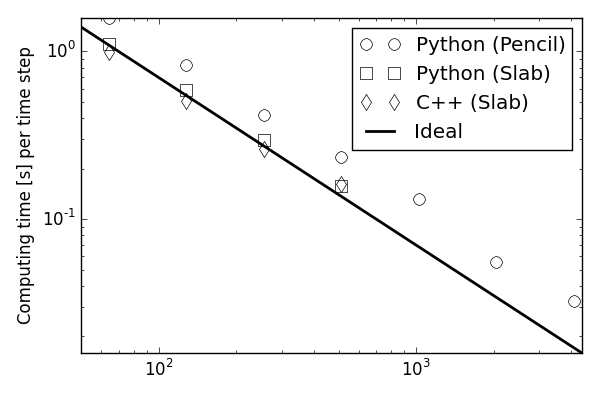}
	\caption{Strong scaling of forward and inverse FFTs. The Python results are 
	non-optimized and C++ is using slab 
		decomposition with MPI communication performed through the FFTW 
		library.}
	\label{fig:strong_FFT_scaling}
\end{figure}

Isolating first the FFTs we plot in Figure
\ref{fig:strong_FFT_scaling} the computing time of one forward and inverse FFT, for both the pencil and slab decompositions and, as for the weak scaling, compare it to a C++ version that is using FFTW. The number of
cores is varied over the range from 64 to 512 for the slab
decomposition, and from 64 to 4096 for the
pencil decomposition. The scaling is very good for both slab and pencil, and good speedup is observed
throughout the entire range. Like for the weak scaling, the Python slab version is seen also here to scale better than the C++ version for the highest number of processors, and even though C++ is approximately 10 \% faster for 64 processors, the Python version is actually faster for 512. The pencil transforms are again seen to be slower than slab, but we achieve good, although not perfect, speedup all the way up to 4096 processors. Note that the results for the pencil transforms depend on the parameter P1, fixed to determine how the two first indices of the computational mesh are distributed. Here we have used $P1 = (8, 16, 16, 32, 32, 64, 64)$ and not optimized any further.

\begin{figure}[ht!]
\subfigure[Scientific Python solver]{
\includegraphics[scale=0.44]{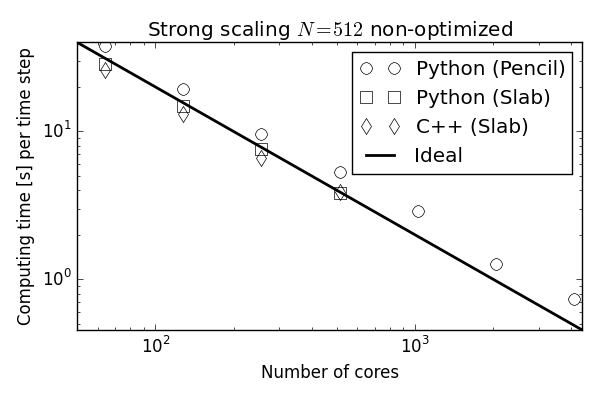}
\label{fig:strong_scaling_shaheen}
}
\subfigure[Python solver optimized with Cython]{
\includegraphics[scale=0.44]{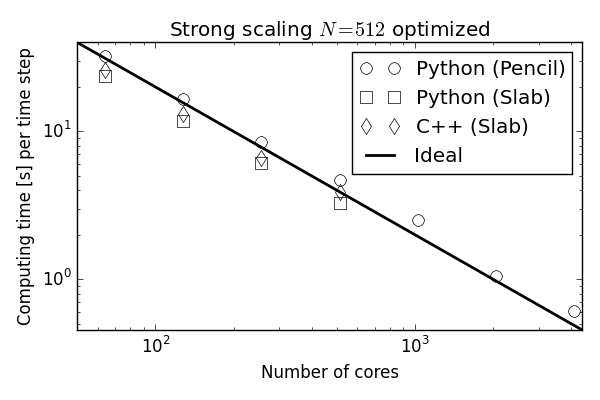}
\label{fig:strong_scaling_shaheen_opt}
}
\caption{Strong scaling of the  DNS solver. The C++ solver uses slab decomposition with MPI communication performed through the FFTW library. The C++ results are the same in \subref{fig:strong_scaling_shaheen} and \subref{fig:strong_scaling_shaheen_opt}.}
\label{fig:Strong_scaling}
\end{figure}

Figure \ref{fig:Strong_scaling} shows the performance of the complete solvers. The outcome is similar to that found in the weak scaling test and the optimized Python/Cython code is seen
to more than match the performance of the C++ solver for this particular
case. Good scaling is observed for the pencil solver all the way up to 4096 processors.

\subsection{Comparison to Tarang}
At this point the reader may still be skeptical since the C++ solver described 
in previous sections  
is not a properly tested and optimized code, and therefore the comparison may  
not seem fair. For this reason we will finally comment on the performance in 
relation to 
the open source C++ pseudo-spectral solver Tarang \cite{tarang}. 
Tarang is 
using the same 4'th order Runge-Kutta integrator as our code, dealiasing is 
performed in the same way and 3D FFT is performed using FFTW with slab 
decomposition. As such, computational cost should be more or less identical and 
the computational speed should be directly comparable to our solver. 
Furthermore, Tarang  has been benchmarked by its own developers on Shaheen. 

With this 
in mind we can collect the timings from Figure 2 of \cite{tarang}, which reads 
that a box of size $1024^3$ (single precision) is running at approximately $50$ 
seconds per time step, and that a perfectly scaled solver would optimally run 
at approximately 32 seconds. From the 
largest mesh used in Figure \ref{fig:Weak_scaling} b) results are plotted for a 
box of size $1024^3$ using 1024 cores. The actual timing reads approximately 18 
seconds, whereas optimal performance 
(perfect weak scaling) would be achieved at slightly less than 15 s. Assuming 
perfect strong 
scaling down to 512 cores, this would correspond to 36/30 seconds for actual 
and optimal performance on the same mesh as used by Tarang. Comparing to 
Tarang, the optimal performance is thus found to be very similar, whereas
the actual is nearly 30 \% faster due to better scaling. This leads us to 
suspect that Tarang has not been using an optimal installation of FFTW on 
Shaheen, because our C++ solver (which is also using FFTW for parallel FFT), is 
scaling much better, and scaling performance depends almost exclusively on the 
3D parallel FFT.

\section{Conclusions}

We have described a parallel Python implementation of an open source 
pseudo-spectral DNS solver for turbulent flows. The solver is implemented using 
standard scientific Python modules (NumPy, MPI for Python) and a very high 
level of 
abstraction, using no more than 100 lines of code, just like a similar MATLAB 
script, but targeting supercomputer platforms. We have then discussed an 
optimization of a small part of this solver through the use of Cython. The 
solver has been verified and benchmarked on the Shaheen supercomputer at the 
KAUST supercomputing laboratory, and we are able to show excellent scaling up 
to several thousand cores with performance matching that of a pure C++ 
implementation.

\section*{Acknowledgements}
M. Mortensen acknowledges support from the 4DSpace Strategic Research 
Initiative at the University of Oslo. M. Mortensen and H. P. Langtangen also 
acknowledge support through a Center of Excellence grant from the Research 
Council of Norway to the Center for Biomedical Computing at Simula Research 
Laboratory. We are also grateful to David Ketcheson and the 
KAUST Supercomputing Laboratory for providing access to Shaheen.

\bibliography{bib.bib}

\begin{thebibliography}{10}

\bibitem{ayala2013}
O.~Ayala and L.~P. Wang.
\newblock {Parallel implementation and scalability analysis of 3D Fast Fourier
  Transform using 2D domain decomposition}.
\newblock {\em Parallel Computing}, 39:58--77, 2013.

\bibitem{petsc-web-page}
S.~Balay, J.~Brown, K.~Buschelman, W.~D. Gropp, D.~Kaushik, M.~G. Knepley,
  L.~Curfman McInnes, B.~F. Smith, and H.~Zhang.
\newblock {PETSc} {W}eb page.
\newblock http://www.mcs.anl.gov/petsc.

\bibitem{brachet1991direct}
M.~E. Brachet.
\newblock Direct simulation of three-dimensional turbulence in the taylor-green
  vortex.
\newblock {\em Fluid dynamics research}, 8(1):1--8, 1991.

\bibitem{canuto1988}
C.~Canuto, M.~Y. Hussaini, A.~Quarteroni, and T.~A. Zang.
\newblock {\em Spectral Methods in Fluid Dynamics}.
\newblock Springer-Verlag New York-Heidelberg-Berlin, 1988.

\bibitem{hit-3d}
S.~G. Chumakov.
\newblock https://code.google.com/p/hit3d/.

\bibitem{cython}
Cython.
\newblock http://cython.org.

\bibitem{deBruynKops15}
S.~de~Bruyn~Kops.
\newblock Classical scaling and intermittency in strongly stratified boussinesq
  turbulence.
\newblock {\em J. Fluid Mechanics}, 775:436--463, 2015.

\bibitem{Ding95}
H.~Q. Ding, R.~D. Ferraro, and D.~B. Gennery.
\newblock A portable {3D FFT} package for distributed-memory parallel
  architectures.
\newblock In {\em {Proceedings of 7th SIAM Conference on Parallel Processing}},
  pages 70--71, 1995.

\bibitem{Enkovaara201117}
J.~Enkovaara, N.~A. Romero, S.~Shende, and J.~J. Mortensen.
\newblock {GPAW} - massively parallel electronic structure calculations with
  python-based software.
\newblock {\em Procedia Computer Science}, 4(0):17 -- 25, 2011.

\bibitem{fenics}
{FEniCS}.
\newblock {http://fenicsproject.org}.

\bibitem{philofluid}
M.~Iovieno~et. al.
\newblock http://areeweb.polito.it/ricerca/ philofluid/.

\bibitem{ketcheson2012}
D.~Ketcheson, K.~Mandli, A.~Ahmadia, A.~Alghamdi, M.~de~Luna, M.~Parsani,
  M.~Knepley, and M.~Emmett.
\newblock Pyclaw: Accessible, extensible, scalable tools for wave propagation
  problems.
\newblock {\em SIAM Journal on Scientific Computing}, 34(4):C210--C231, 2012.

\bibitem{mpi_import}
A.~Langton.
\newblock {https://github.com/langton/MPI$\_$Import}.

\bibitem{Lee2013}
M.~Lee, N.~Malaya, and R.~D. Moser.
\newblock Petascale direct numerical simulation of turbulent channel flow on up
  to 786k cores.
\newblock In {\em Proceedings of the International Conference on High
  Performance Computing, Networking, Storage and Analysis}, pages 61:1--61:11,
  New York, USA, 2013.

\bibitem{Li2010}
N.~Li and S.~Laizet.
\newblock {2DECOMP\&FFT - A highly scalable 2D decomposition library and FFT
  interface}.
\newblock In {\em Cray User Group conference, Edinburgh}. 2010.

\bibitem{gpaw05}
J.~J. Mortensen, L.~B. Hansen, and K.~W. Jacobsen.
\newblock Real-space grid implementation of the projector augmented wave
  method.
\newblock {\em Phys. Rev. B}, 71(3):035109, 2005.

\bibitem{Mortensen2015}
M.~Mortensen and K.~Valen-Sendstad.
\newblock {Oasis: A high-level/high-performance open source Navier - Stokes
  solver}.
\newblock {\em Computer Physics Communications}, 188(0):177 -- 188, 2015.

\bibitem{mpi4py}
{MPI for Python}.
\newblock https://bitbucket.org/mpi4py/mpi4py/.

\bibitem{numba}
Numba.
\newblock http://numba.pydata.org.

\bibitem{numexpr}
Numexpr.
\newblock http://github.com/pydata/numexpr.

\bibitem{numpy}
NumPy.
\newblock http://www.numpy.org.

\bibitem{orzag71}
S.~A. Orzag.
\newblock {On the Elimination of Aliasing in Finite Difference Schemes by
  Filtering High-Wavenumber Components}.
\newblock {\em {Journal of the Atmospheric Sciences}}, 28:1074, 1971.

\bibitem{p3dfft}
D.~Pekurovsky.
\newblock {P3DFFT} http://code.google.com/p/ p3dfft/.

\bibitem{pekurovsky2012}
D.~Pekurovsky.
\newblock {P3DFFT}: a framework for parallel computations of {F}ourier
  transforms in three dimensions.
\newblock {\em {SIAM Journal on Scientific Computing}}, 34(4), 2012.

\bibitem{Pi13}
M.~Pippig.
\newblock {PFFT} - {A}n extension of {FFTW} to massively parallel
  architectures.
\newblock {\em SIAM J. Sci. Comput.}, 35:C213 -- C236, 2013.

\bibitem{PlimptonFFT}
S.~J. Plimpton.
\newblock {Parallel FFT Subroutine Library
  http://www.sandia.gov/$\sim$sjplimp/docs/fft/ README.html}.

\bibitem{scalablepython}
{Scalable Python}.
\newblock {https://github.com/CSC-IT-Center-for-Science/scalable-python}.

\bibitem{turbo}
B.~Teaca.
\newblock http://aqua.ulb.ac.be/home/turbo/.

\bibitem{van2011numpy}
Stefan Van Der~Walt, S~Chris Colbert, and Gael Varoquaux.
\newblock {The NumPy array: a structure for efficient numerical computation}.
\newblock {\em Computing in Science \& Engineering}, 13(2):22--30, 2011.

\bibitem{tarang}
M.~Vermi.
\newblock http://turbulence.phy.iitk.ac.in/doku.php.

\bibitem{weave}
Weave.
\newblock http://docs.scipy.org/doc/scipy/reference/ tutorial/weave.html.

\end{thebibliography}

\section*{Appendix}
\label{sec:appendix}
In this Appendix we show a complete Python script that solves Eq. 
(\ref{eq:NSfinal}) with slab decomposition and a fourth order Runge Kutta 
method for the Taylor Green test case. Saved as \inpyth{TG.py} it runs in 
parallel on 128 processors with command \inpyth{mpirun -np 128 python TG.py}.

\newpage
\onecolumn
\begin{python_num}
from numpy import *
from numpy.fft import fftfreq, fft, ifft, irfft2, rfft2
from mpi4py import MPI

nu = 0.000625
T = 0.1
dt = 0.01
N = 2**7
comm = MPI.COMM_WORLD
num_processes = comm.Get_size()
rank = comm.Get_rank()
Np = N / num_processes
X = mgrid[rank*Np:(rank+1)*Np, :N, :N].astype(float)*2*pi/N
U     = empty((3, Np, N, N))
U_hat = empty((3, N, Np, N/2+1), dtype=complex)
P     = empty((Np, N, N))
P_hat = empty((N, Np, N/2+1), dtype=complex)
U_hat0  = empty((3, N, Np, N/2+1), dtype=complex)
U_hat1  = empty((3, N, Np, N/2+1), dtype=complex)
dU      = empty((3, N, Np, N/2+1), dtype=complex)
Uc_hat  = empty((N, Np, N/2+1), dtype=complex)
Uc_hatT = empty((Np, N, N/2+1), dtype=complex)
U_mpi   = empty((num_processes, Np, Np, N/2+1), dtype=complex)
curl    = empty((3, Np, N, N))
kx = fftfreq(N, 1./N)
kz = kx[:(N/2+1)].copy(); kz[-1] *= -1
K = array(meshgrid(kx, kx[rank*Np:(rank+1)*Np], kz, indexing='ij'), dtype=int)
K2 = sum(K*K, 0, dtype=int)
K_over_K2 = K.astype(float) / where(K2 == 0, 1, K2).astype(float)
kmax_dealias = 2./3.*(N/2+1)
dealias = array((abs(K[0]) < kmax_dealias)*(abs(K[1]) < kmax_dealias)*
                (abs(K[2]) < kmax_dealias), dtype=bool)
a = [1./6., 1./3., 1./3., 1./6.]
b = [0.5, 0.5, 1.]

def ifftn_mpi(fu, u):
    Uc_hat[:] = ifft(fu, axis=0)
    comm.Alltoall([Uc_hat, MPI.DOUBLE_COMPLEX], [U_mpi, MPI.DOUBLE_COMPLEX])
    Uc_hatT[:] = rollaxis(U_mpi, 1).reshape(Uc_hatT.shape)
    u[:] = irfft2(Uc_hatT, axes=(1, 2))
    return u

def fftn_mpi(u, fu):
    Uc_hatT[:] = rfft2(u, axes=(1,2))
    U_mpi[:] = rollaxis(Uc_hatT.reshape(Np, num_processes, Np, N/2+1), 1)
    comm.Alltoall([U_mpi, MPI.DOUBLE_COMPLEX], [fu, MPI.DOUBLE_COMPLEX])
    fu[:] = fft(fu, axis=0)
    return fu

def Cross(a, b, c):
    c[0] = fftn_mpi(a[1]*b[2]-a[2]*b[1], c[0])
    c[1] = fftn_mpi(a[2]*b[0]-a[0]*b[2], c[1])
    c[2] = fftn_mpi(a[0]*b[1]-a[1]*b[0], c[2])
    return c

def Curl(a, c):
    c[2] = ifftn_mpi(1j*(K[0]*a[1]-K[1]*a[0]), c[2])
    c[1] = ifftn_mpi(1j*(K[2]*a[0]-K[0]*a[2]), c[1])
    c[0] = ifftn_mpi(1j*(K[1]*a[2]-K[2]*a[1]), c[0])
    return c

def computeRHS(dU, rk):
	if rk > 0:
		for i in range(3):
			U[i] = ifftn_mpi(U_hat[i], U[i])
	curl[:] = Curl(U_hat, curl)
	dU = Cross(U, curl, dU)
	dU *= dealias
	P_hat[:] = sum(dU*K_over_K2, 0, out=P_hat)
	dU -= P_hat*K
	dU -= nu*K2*U_hat
	return dU

U[0] = sin(X[0])*cos(X[1])*cos(X[2])
U[1] =-cos(X[0])*sin(X[1])*cos(X[2])
U[2] = 0
for i in range(3):
U_hat[i] = fftn_mpi(U[i], U_hat[i])

t = 0.0
tstep = 0
while t < T-1e-8:
	t += dt; tstep += 1
	U_hat1[:] = U_hat0[:] = U_hat
	for rk in range(4):
		dU = computeRHS(dU, rk)
		if rk < 3: U_hat[:] = U_hat0 + b[rk]*dt*dU
		U_hat1[:] += a[rk]*dt*dU
	U_hat[:] = U_hat1[:]
	for i in range(3):
		U[i] = ifftn_mpi(U_hat[i], U[i])

k = comm.reduce(0.5*sum(U*U)*(1./N)**3)
if rank == 0:
    assert round(k - 0.124953117517, 7) == 0

\end{python_num}

\end{document}